\documentclass[journal=jacsat,manuscript=article]{achemso}

\usepackage[version=3]{mhchem} 

\usepackage{bm}

\usepackage{xcolor}
\newcommand{\cblue }{\color{black}}

\author{Arindam Panda} 
\altaffiliation{}
 \email{arindam19@iiserb.ac.in}
\author{Sunil P. Singh}
\email{spsingh@iiserb.ac.in}
\affiliation[Indian Institute of Science Education and Research]
{Department of Physics, Indian Institute of Science Education and Research, \\ Bhopal 462 066, Madhya Pradesh, India}

\title[An \textsf{achemso} demo]
  {Folding-unfolding transition  of  active polymer on the reconfiguration of bidirectional tangential active force}

\abbreviations{ABP, APLP,LJ}
\keywords{American Chemical Society, \LaTeX}

\begin{document}

\begin{tocentry}
    \centering
         A schematic of a bidirectional active polymer, where the arrows display the direction of active force.  A pair of half-headed arrows illustrates the polarity reversal of the active force.  
         \includegraphics[width=2.75cm]{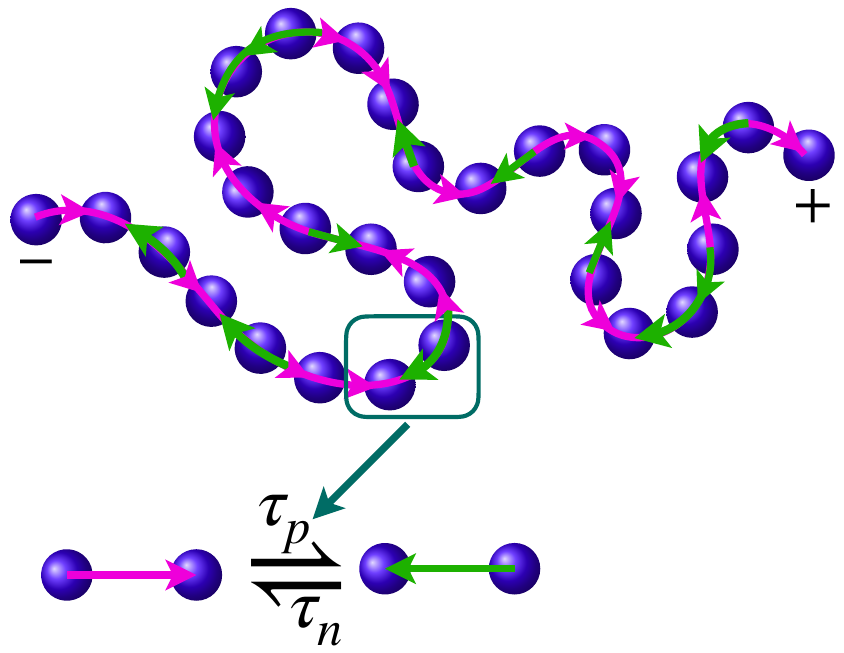}        
 
\end{tocentry}

\begin{abstract}
   The role of active stress on the conformational dynamics of a polymer has drawn significant interest due to its potential applications in understanding the energy landscape of protein structures, buckling of biopolymers,  genomic spatial organization and their large-scale coherent dynamics. We present a model of bidirectional active force that acts along the polymer's tangent, with its direction stochastically reversing between head-to-tail and tail-to-head orientations. The active polymer shows a structural transition from a random coil-like state to a compressed state with variations in the active force, directional (polarity) reversal rate, and their fraction. Furthermore, the polymer re-swells and stretches more than its passive limit for a large active force.  The polymer’s radius of gyration follows the ideal chain-like scaling relation, $R_g^2\sim N_m^{2\nu}$ with an exponent $\nu\approx 1/2$,  in both the compressed and swelled states. The bidirectional active force also drives dynamical transitions, where the effective diffusivity abruptly shifts from a linear to quadratic increase. Similarly,  in the regime of large activity, the linear decrease of the longest relaxation time of the polymer changes behavior to a power-law behavior $\mathrm{Pe}^{-4/3}$ with P\'eclet number.  We have shown that the active polymer's conformational, relaxation, and diffusive behaviors display a transition from an active polar linear polymer model (APLP) to an active Brownian particle (ABP) polymer model with the increase in the fraction of the opposite polarity and their reconfiguration time. 
\end{abstract}

\section{Introduction}
The folding-unfolding transition of polymer conformations driven by the interplay of entropic, energetic, and external fields is widely studied subject across the discipline \cite{shakhnovich2006protein,rubinstein2003polymer,onuchic1997theory,england2008potential,england2011allostery,perunov2014quantitative,Jing_PNAS_2011,Zhou_PRL_2011,radhakrishnan2021collapse,alexander2008dynamics,harasim2013direct}. 
In the past few years, there has been growing interest in understanding the conformational transitions of a polymer under the influence of activated processes. These processes, coupled with biopolymers, perform various important biological functions,  particularly the compaction of chromatin, the buckling of microtubules and actin filaments,  and the hierarchical structural organization of proteins and chromatin\cite{misteli2020self,goychuk2023polymer,davidson2019dna,goychuk2024delayed}.  Such compaction may lead to inhomogeneous folding and induce effective attraction between distant segments of a polymer\cite{goychuk2023genom}.
Moreover, active processes are vital for a wide range of cellular activities\cite{elgeti2015physics,bechinger2016active,winkler2020physics}, {\it viz.}   vesicle's transport \cite{ahmed2014active,raote2019protein,goode2000functional,schuh2011actin,kamal2000connecting}, microswimmer's motility\cite{elgeti2015physics,huang2019adaptive},   and the replication, packaging\cite{patton2000genome}  and segregation of genetic materials\cite{potten1978segregation,gordon2000dna}.

Theoretical models have primarily focused on the uniform distribution of active force, either unidirectionally quenched along the polymer's conformation\cite{chelakkot2014flagellar,winkler2020physics,duman2018collective,anand2018structure,isele2015self,elgeti2015physics,mokhtari2019dynamics,wang2015one,duman2018collective,prathyusha2018dynamically} or a randomly oriented colored active noise on each monomer \cite{anand2020conformation,eisenstecken2016conformational,das2019deviations,kaiser2015does}. The active force quenched along the polymer's conformation is referred to as a tangentially driven active polar linear polymer (APLP)\cite{chelakkot2014flagellar,winkler2020physics,duman2018collective,anand2018structure,isele2015self,elgeti2015physics,mokhtari2019dynamics,wang2015one,duman2018collective,prathyusha2018dynamically}. 
A key feature of the APLP model is that the unidirectional tangential active force renders large-scale compression even in the absence of any attraction among the monomers\cite{philipps2022tangentially,anand2018structure,bianco2018globulelike}. 
The latter model, where each polymeric monomers are subjected to colored active noise whose direction randomly changes called the active Brownian particle polymer model (ABP)\cite{das2019deviations,anand2020conformation,martin2019active,liao2020extensions,mousavi2019active,ghosh2014dynamics,panda2023characteristic}. 
In contrast, the ABP model can lead to a subtle compression followed by large-scale swelling of the polymer\cite{das2019deviations,mokhtari2019dynamics,wang2015one,duman2018collective,prathyusha2018dynamically,anand2020conformation,martin2019active,liao2020extensions}. 

The bidirectional transport of lipid droplets, vesicles, adenoviruses, granules, mitochondria, and secretory vesicles along cytoskeleton or axonemal microtubule tracks in cells, neurons, cilia, and flagella is a well-established phenomenon\cite{gross2004hither,welte2004bidirectional,hancock2014bidirectional,chevalier2006axonal,d2023vesicles}. 
Such transports are facilitated by motor proteins, primarily kinesin and dynein,  where kinesin moves toward the positive end of the microtubule. On the other hand, dynein motor proteins move toward the negative end\cite{d2023vesicles}. It has been observed that the dynactin, a motor protein complex, can facilitate bidirectional transport\cite{deacon2003dynactin,ross2006processive,Derewenda2007}.
Additionally,  dynein and kinesin can engage in bidirectional transport in the presence of certain regulatory proteins \cite{ross2006processive,hancock2014bidirectional, Egan2012}.
There are numerous observations indicating that the motor proteins impose bidirectional active force along the filament tracks\cite{hirokawa2005molecular,zidovska2013micron,goloubinoff2018chaperones,davidson2019dna,saintillan2018extensile,putpre2019}.  
These bidirectional forces may contribute to various complex dynamical behaviors, potentially giving rise to intriguing phenomena such as vortex formation, spiral patterns, and swirling motions. At the macroscopic scale, the stochastic reversal of the direction of motion of the various microorganisms, notably {\it Myxococcus xanthus, Pseudomonas putida, Pseudoalteromonas haloplanktis,  and Pseudomonas citronellolis} is an established phenomenon \cite{santra2021active,wu2009periodic,liu2019self,barbara2003bacterial,tamulonis2011modeling,abbaspour2023effects}.

\begin{figure}[t]
    \centering
    \includegraphics[width=\columnwidth]{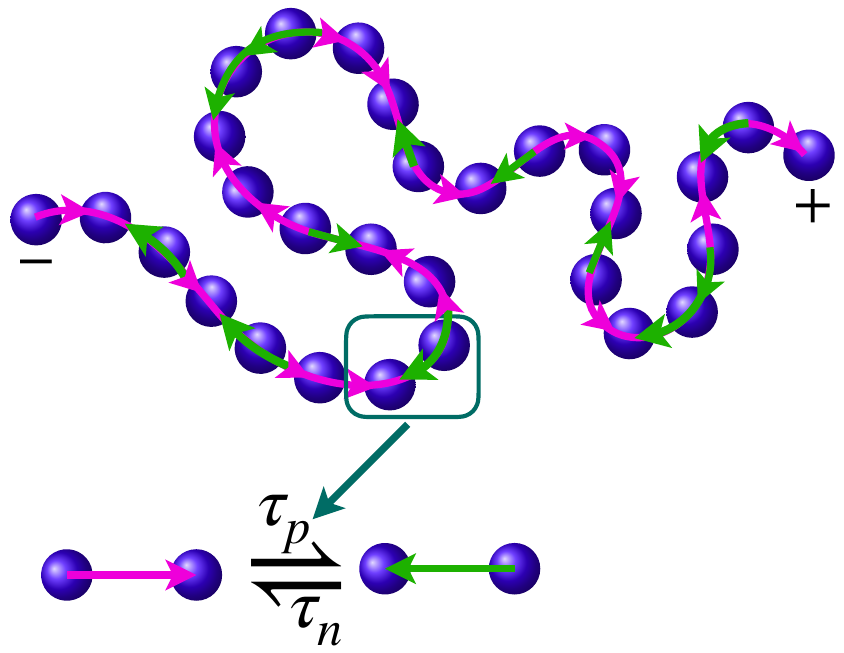}       
    \caption{ A schematic of the bidirectional active linear polymer. The arrows display the direction of the active force on each bond.  A pair of half-headed arrows illustrates the direction reversal of active force on each bond attempted stochastically at a time interval $\tau_n$.    }
    \label{Fig:schematic}
\end{figure}

Understanding a polymer's response to bidirectional tangential active force is crucial; however, this aspect remains unexplored. 
We present a simulation model of a flexible linear polymer that considers active force along each bond. The polarity of this active force can be directed either toward the head monomer  (+)  or the tail monomer (-), for an illustration, see Fig.~\ref{Fig:schematic}. 
The fraction of positively and negatively oriented active bonds and the polarity reversal rates are also considered variables in the model. 
The polarity reversal rates emulate the effect of attachment and detachment of active agents\cite{peruani2012collective,mcbride2001bacterial,tamulonis2011modeling,turner2000real,gupta2019morphological,mahajan2022self}.
This can be modeled using a Poisson-distributed random noise. Figure~\ref{Fig:schematic} illustrates the polarity of the active force and its reversal, along with the corresponding signs of the head (+) and tail (-) monomers.

Our simulations unveil that nearly equal fractions of positive and negative polarity active forces instigate a conformational transition. Here, a compressed polymer, for the case of unidirectional tangential active force, unfolds to a swollen state. This swelling is further enhanced by parameters such as the strength of the active force, the fraction of the negative polarity bonds, and the polarity reversal rate. 
We demonstrate a transition in the polymer's dynamical and relaxation behavior from the APLP to the ABP polymer model as the polarity reversal rate and the negative active bonds vary. We have analyzed the structural transition in terms of physical quantities, bond-bond correlations, contact frequency of monomers, and inhomogeneous stretching under bidirectional active force.

The manuscript is organized as follows: a detailed description of the model is presented in the "simulation model" section.   The main simulation results are presented in subsections named conformations, contact probability distribution,  segmental stretching, bond-bond correlations, relaxation time,  effective diffusivity of the polymer, and stochastic polarity reversal. Results are summarized in the conclusion Section. 

\section{Simulation Model} 
We model a flexible linear polymer with $N_m$ active monomers, where monomers are linearly linked to their adjacent ones by a harmonic potential,
\begin{equation}
{U_s}= \frac{\kappa_s}{2} \sum_{i=1}^{N_m-1} (|{\bm r}_{i+1} - {\bm r}_i| - \ell_0)^2,
\label{Eq:spring}   
\end{equation}
where $\kappa_s$, $\bm{r}_i$, $\ell_0$   are the spring constant,  the position vector of the $i^{\mathrm{th}}$ monomer, and the equilibrium bond length, respectively. Additionally, excluded-volume interactions among
chain monomers are incorporated by the repulsive, truncated, and shifted Lennard-Jones (LJ) potential ($U_{LJ}$), given as
\begin{equation}
    U_{LJ} = 4 \epsilon\sum_{i>j}^{N_m} \left[\left(\frac{\sigma}{r_{ij}}\right)^{12} -\left(\frac{\sigma}{r_{ij}}\right)^{6} + \frac{1}{4}\right], r_{ij} \leq r_c
\end{equation} 
where $r_c$ is the cut-off distance between the monomers for $r_c \geq 2^{1/6} \sigma$, $U_{LJ} = 0$. Here, $\sigma$ is the LJ diameter of the monomers, $\epsilon$ is their interaction energy, and ${r}_{ij} = |\bm{r}_j-\bm{r}_i|$ is the distance between  $i^{\mathrm{th}}$  and $j^{\mathrm{th}}$ monomers. 

The total active force on the polymer is given here as
\begin{equation}
{\bm F}_a = \sum_{i=1}^{N_m - 1}p_i f_a {\hat {\bm t}}_i,
\end{equation}
where $f_a$ denotes the active force strength per unit bond length, ${\hat{\bm t}_i} = (\bm{r}_{i+1} - \bm{r}_i) / \left | \bm{r}_{i+1} - \bm{r}_i \right|$ represents the $i^{\mathrm{th}}$ tangent vector of the polymer, and $p_i$ indicates the polarity of the $i^{\mathrm{th}}$ bond.  The polarity has two possible values: $p_i = +1$ indicates activity towards the head monomer, while $p_i = -1$ denotes activity directed towards the tail. 
 The fraction of negative polarity is expressed as $x_{n}= \frac{N_{p^{(-)}}}{N_m-1}$, along with $N_m-1=N_{p^{(+)}}+N_{p^{(-)}}$,  $N_{p^{(+)}}$ and $N_{p^{(-)}}$ are the total number of the active bonds with positive and negative polarities.  

The direction of the active force on a given bond ($p_i \hat {\bm t}_i$) stochastically reverses from head-to-tail or tail-to-head with reversal rates $\lambda_p$, $\lambda_n$, respectively. This breaks the symmetry and can introduce unequal fractions of positive and negative active bonds, for the case of $\lambda_p\neq \lambda_n$. The reversal rate of the polarity can be expressed as follows ${(p_i =-1) \rightleftharpoons  (p_i=1)}$ at every  $\frac{1}{\lambda_n}$  time. The polarity reversal of negative bonds is attempted with a probability of $x_n=\frac{\lambda_n}{\lambda_n + \lambda_p}$, and positive bonds with $1-x_n$.  Importantly, it is worth noting here that this manuscript considers all results for a fixed polarity reversal rate at an interval $\tau_n=\frac{1}{\lambda_n}$. 

Additionally, we have also considered that the waiting time for the two consecutive polarity reversal events chosen from a Poisson-distributed random number with the average reversal time $\frac{1}{\lambda_{n}}$ and $\frac{1}{\lambda_{p}}$; particularly, these results are presented in the subsection named stochastic polarity reversal. 

The  equations of motion of the monomers in the overdamped limit are given by
\begin{equation}
    \zeta {\dot{\bm r}}_i(t) = \bm{F}_i(t)  + \bm{F}_i^a(t) + \bm{\Gamma}_i(t),
\end{equation}
where $\zeta$ is the drag coefficient,  $\bm{F_i}(t)$ includes bond and excluded-volume interactions. Further, $\bm{\Gamma}_i$ is a Gaussian white noise with the first moment $\langle \bm{\Gamma}_i(t) \rangle = 0$  and the correlations  $\langle \bm{\Gamma}_i(t) \cdot \bm{\Gamma}_j(t') \rangle = 6k_B T \zeta \delta_{ij} \delta(t-t')$\cite{risken1989fokker-planck,doi1988theory}. It describes thermal noise, where $k_B$ is Boltzmann's constant and $T$ is the bath temperature. {\cblue  
The active force on the $i^{\mathrm{th}}$ monomer is given by, 
$    \mathbf{F}_i^a =  \frac{f_a}{2} \left(p_{i-1} {\hat {\bm t}}_{i-1}+ p_{i}{\hat {\bm t}}_{i}\right)$.  The active force on the terminal ends $\mathbf{F}_1^a= \frac{f_a}{2} \left(p_1 \hat{\bm t}_1\right)  $ and $\mathbf{F}_{N_m}^a= \frac{f_a}{2} \left(p_{N_m-1} \hat{\bm t}_{N_m-1}\right)  $.

The detailed flow chart of the simulation's algorithm is provided in  SI-Fig. S5\cite{supp,panda_2025_17011905}. }

\subsection{Parameters}
All simulation parameters are in reduced units of the  LJ diameter $\sigma$, thermal energy $k_B T$, and a characteristic time scale $\tau$ defined as $\tau = \sigma^2 /D_t$, where $D_t = k_B T / \zeta$ is the translational diffusion coefficient of a free monomer.  Forces are scaled in units of $k_B T / \sigma$.  
The spring constant $\kappa_s$ is chosen from $(1$ - $2.5) \times 10^4 k_B T / \sigma^2$ to avoid bond stretching under large active force conditions. The equilibrium bond length is $\ell_0 = \sigma$, and the LJ energy parameter is fixed at $\epsilon / k_B T = 1$.  The strength of activity in terms of the dimensionless P\'eclet number is defined as:  
$\mathrm{Pe} = \frac{f_a \ell_0}{k_B T}$, where $\mathrm{Pe}$ is varied between $(0 - 200)$. 
The simulations are performed with a time step $\Delta t$ in the range $(10^{-5} - 10^{-4})\tau$. {\cblue To avoid simulation artifacts, for higher activity, we use smaller time steps, which prevents overstretching of the bonds.} The reversal time $\tau_n$ varied in the range $(10^{-3} - 10^2)\tau$. Furthermore, to establish the scaling relations, we consider polymers of chain lengths $N_m = 100, 200$, $400$, and $800$, corresponding to contour lengths $L = 99 \ell_0, 199\ell_0$, $399\ell_0$, and $799\ell_0$, respectively.

\section{Results and discussion}

\subsection{Conformations}
The role of unidirectional tangential active force on the characteristic features of flexible and semi-flexible polymers' conformation and their tendency of compression is revealed in the previous studies\cite{bianco2018globulelike,foglino2019non,anand2018structure}.
This study focuses on the structural behavior of bidirectional active polymer as a function of the fraction of negative side-directed bond $x_n$,  P\'eclet number $\mathrm{Pe}$, and polarity reversal rate $\lambda_n$. The conformational change is estimated in terms of the average mean-square radius of gyration of polymer defined as $R_g^2 =  \frac{1}{N_m} \sum_{i=1}^{N_m}\langle(\bm{r}_i - \bm{r}_{cm})^2 \rangle$, with $\bm{r}_{cm}$ denoting the center-of-mass of the polymer. {\cblue The radius of gyration $R_g^2 $ describes the average size of a polymer and the spatial distribution of a polymer’s monomers around its center of mass. The $R_{g_0}^2$ is defined here to represent the radius of gyration of a passive polymer in equilibrium. The active force drives the polymer out of equilibrium; therefore, comparing the structural change of the polymer with respect to its equilibrium value $ R^2_{g_0} $ reveals the relative change under the influence of activity. A ratio $ {R^2_g}/{R^2_{g_0}} < 1$
indicates compression or folding due to activity, while 
$ {R^2_g}/{R^2_{g_0}} > 1$
suggests swelling or unfolding of the polymer.
In this work, we consider good solvent conditions; therefore, the polymer’s equilibrium radius of gyration scales with chain length, $R_{g_0} \sim N_m^{\nu}$
with $\nu = 3/5$.  On the other hand, the scaling exponent is $\nu =1/2$ in a theta solvent.} 

\begin{figure}[t]
    \includegraphics[width=0.475\columnwidth]{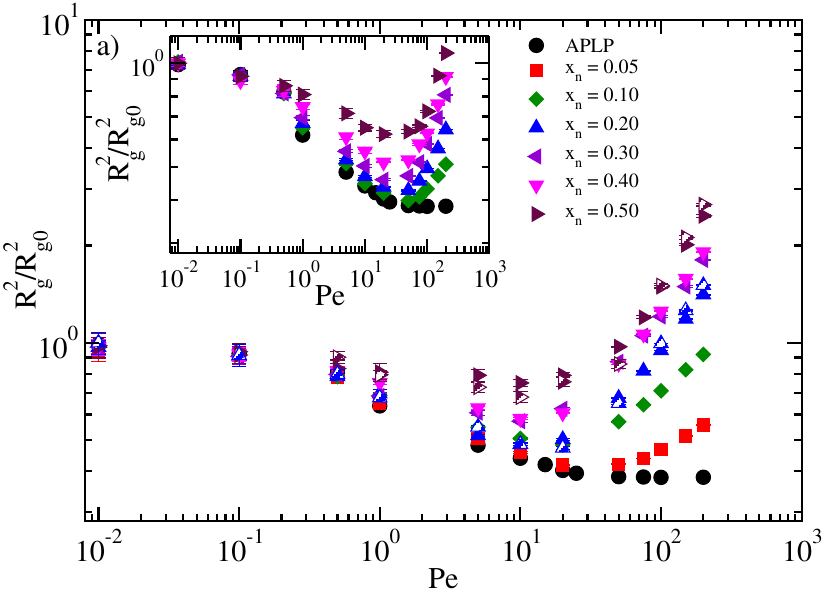}
       \includegraphics[width=0.475\columnwidth]{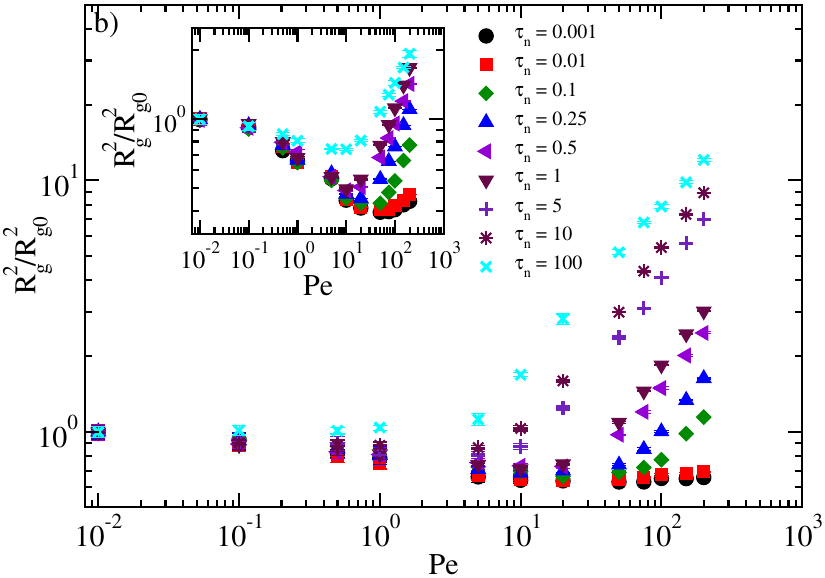}
    \caption{a) {\cblue The relative change in the mean-square radius of gyration  $R_g^2 / R_{g_0}^2$ of the active polymer with respect to its equilibrium value  $ R_{g_0}^2$ as a function of P\'eclet number $\mathrm{Pe}$. The data is presented for different values of $x_n$, as indicated in the plot, at a polarity reversal time $\tau_n = 0.5$.
  Inset captures the variation of $R_g^2/R_{g_0}^2$  as a function of   $\mathrm{Pe}$  at  $\tau_n=0.1$ while keeping other parameters same as the main plot.  b)  The  variation of $R_g^2 / R_{g_0}^2$, as a function of    $\mathrm{Pe}$. The main plot shows the transition from the compression to stretching of the polymer as the reversal time $\tau_n$  is increased at equal polarity fraction $x_n = 0.5$. The inset presents the corresponding behavior at $x_n = 0.2$. 
 Note that the shaded symbols represent the case of a chain length of $N_m = 400$, while the filled one corresponds to $N_m=200$. }}
    \label{Fig:percentage}
\end{figure}

The computed radius of gyration is presented in  Fig.~\ref{Fig:percentage}. 
In the unidirectional (APLP) model, the polymer exhibits significant compression with increasing activity ($\mathrm{Pe}$) compared to its passive counterpart.\cite{bianco2018globulelike,foglino2019non,anand2018structure} As the bidirectionality is introduced by randomly placing the active force of negative polarity results in a distinctive non-monotonic trend. In addition to compression in the intermediate regime, the polymer also attains swelling in the limit of large $\mathrm{Pe}$. Figure~\ref{Fig:percentage}-(a) distinctly illustrates that the polymer undergoes significant swelling with an increase in the fraction of active bonds of negative polarity for large P\'eclet numbers ($\mathrm{Pe}$) at polarity reversal time $\tau_n=0.5$.  Notably, compression of the polymer is significantly reduced with increasing  $x_n$.  This behavior is thoroughly investigated for various $\tau_n$ and $x_n$; as the inset of Fig.~\ref{Fig:percentage}-(a) presents results at $\tau_n = 0.1$.  For  $\tau_n=0.1$, the polymer's structural behavior nearly retains all the qualitative features of the main plot.

Now, we examine the role of polarity reversal time $\tau_n$ on the radius of gyration while maintaining a fixed fraction of opposite polarity of active force onto the polymer. The variation of reversal rate while keeping a fixed fraction  $x_n$ is equivalent to stating that the reversal time of positive polarity ($\tau_p$) is also varied along with $\tau_n$ to maintain the same fraction $x_n$.
For $\tau_n<<1$,  polarity reversal has a negligible impact on the structure of the polymer, as   Fig.~\ref{Fig:percentage} (b) shows  $R^2_g/R^2_{g0}$  for $\tau_n=0.001$ and $0.01$ overlap with the curve of APLP (black curve).
On further increasing  $\tau_n$, the scale of compression reduces and the location of maximum compression shifts to the lower values of $\mathrm{Pe}$,  as Fig.~\ref{Fig:percentage} (b) illustrates.  Additionally, for $\tau_n\geq 0.1$, the radius of gyration once again displays the non-monotonic behavior where compression is followed by stretching of the chain in the limit of large $\mathrm{Pe}$.  More importantly, as reversal time increases, $\tau_n\gg 1$, the compression of the polymer disappears, and the gyration radius increases monotonically with increasing {\cblue P\'eclet} number $\mathrm{Pe}$ at $x_n=0.5$.  A similar qualitative behavior at $x_n=0.2$ is illustrated in the inset of Fig.~\ref{Fig:percentage}-(b). In summary, with increasing $\tau_n$, compression is suppressed, and in the limit of large $\tau_n$, it may disappear.  

\begin{figure}[t]
    \centering
    \includegraphics[width=0.475\textwidth]{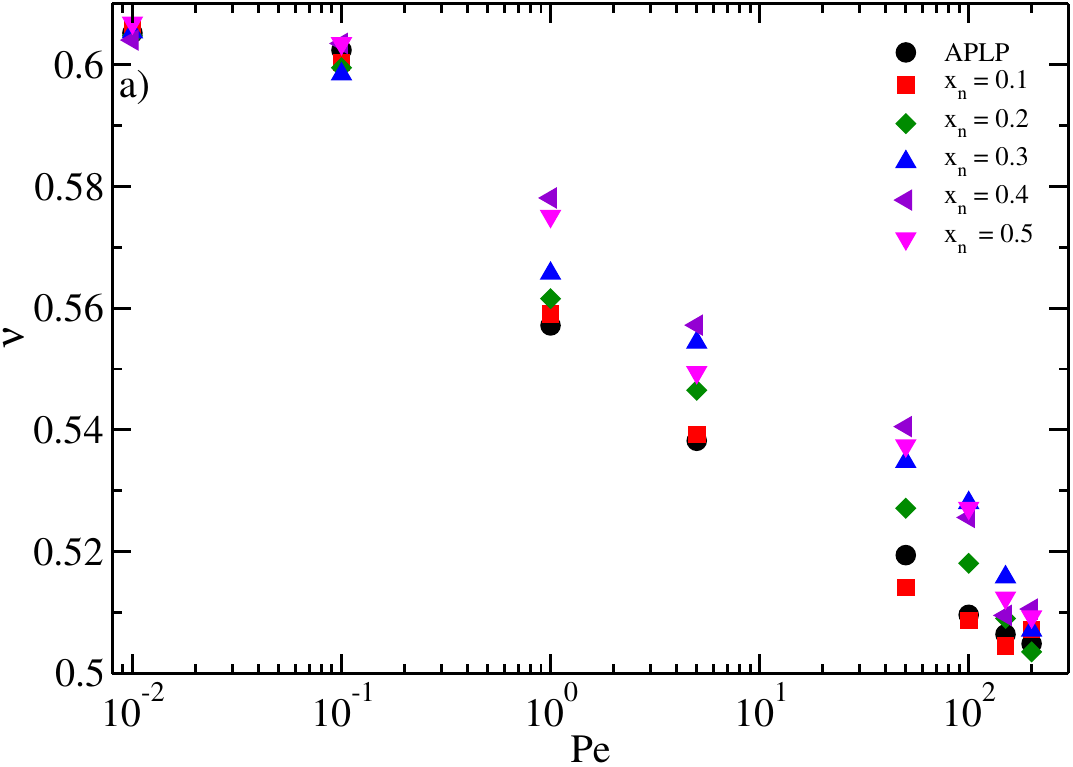}
    \includegraphics[width=0.475\columnwidth]{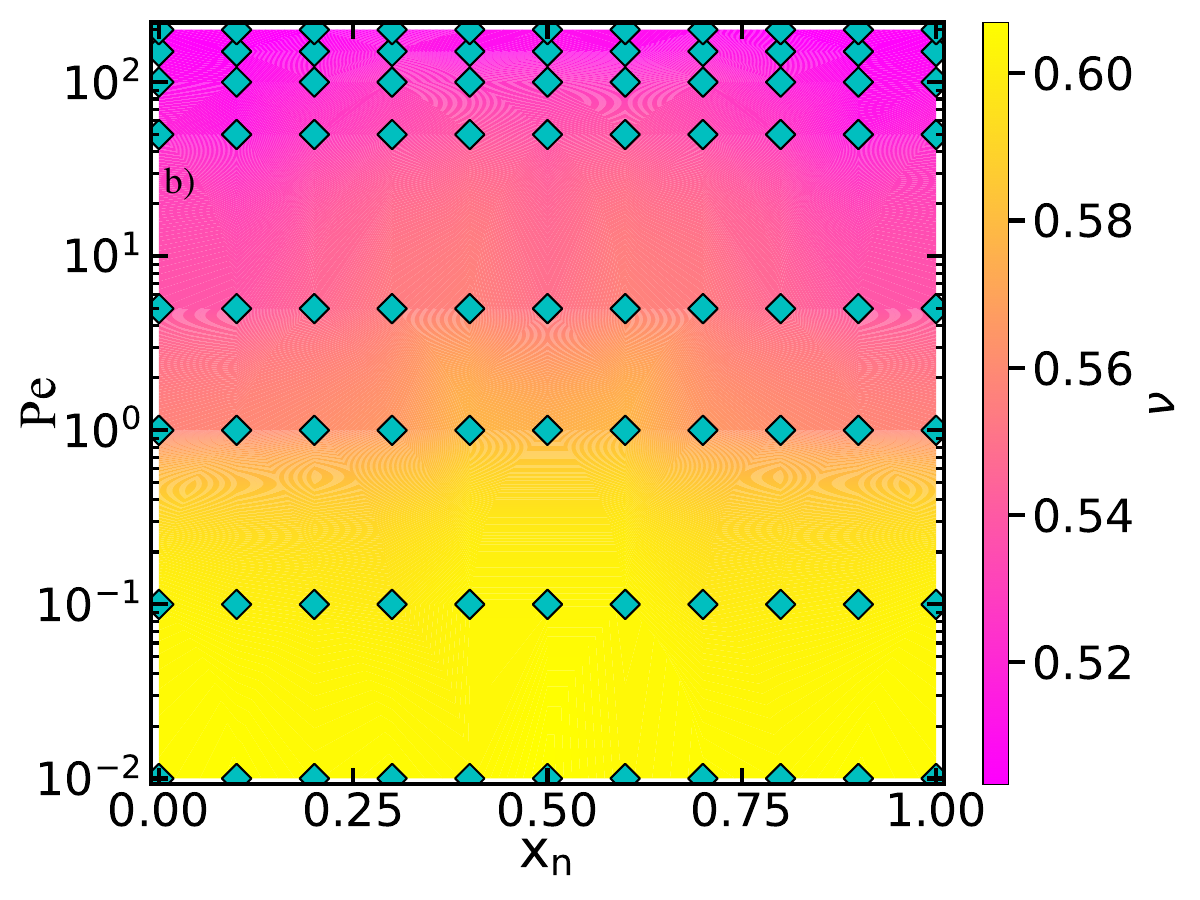}       
    \caption{ \cblue a)  The variation of the scaling exponent $\nu$ as a function of the P\'eclet number  for various  $x_n$.  b) The colormap on the $\mathrm{Pe}$–$x_n$ plane illustrates how the exponent $\nu$ changes across parameter space. The color bar and values indicate a smooth transition from $\nu \approx 3/5$ (self-avoiding chain) to $\nu \approx 1/2$ (an ideal chain). For a symmetric graphical presentation, the x-axis of the colormap $x_n$ is plotted in the range from $0$(APLP) to $1.0$.}
    \label{Fig:colormap}
\end{figure}

The stretching behavior of the bidirectional active polymer with increasing $x_n$ and $\tau_n$ can be explained in terms of local polymer stretching, as discussed in later sections.

 Now, we probe how the scaling relation of a polymer is modified in compressed and stretched states due to non-thermal active force. For this, we compute the radius of gyration of the polymer for various chain lengths, which exhibits a power-law relation expressed as  $R_g^2 \sim N_m^{2\nu}$ in the considered range of $N_m=100,200,400$, and $800$. Figure~\ref{Fig:colormap}-a  presents the polymer's scaling exponent $\nu$  as a function of $\mathrm{Pe}$ for various $x_n$ at a  given $\tau_n=0.5$.  The scaling exponent monotonically decreases from $3/5$ to $1/2$ with  $\mathrm{Pe}$. These values are further presented in $x_n$-$\mathrm{Pe}$ plane as a colormap with a symmetry about $x_n=0.5$. 
 Figure~\ref{Fig:colormap}  reveals an ideal chain-like scaling exponent for all $x_n$ in the limit of large $\mathrm{Pe}>1$. 
 In summary,  a transition from a passive coil-like polymer's exponent  $\nu\approx 3/5$  to the ideal chain's exponent  $\nu\approx1/2$ in the limit of large activity is observed, despite only repulsive interactions among monomers. This transition is purely driven by the activity that dominates over the excluded volume interactions, leading to the ideal chain-like features.   

\begin{figure}[h!]
\centering
	\includegraphics[width=\textwidth]{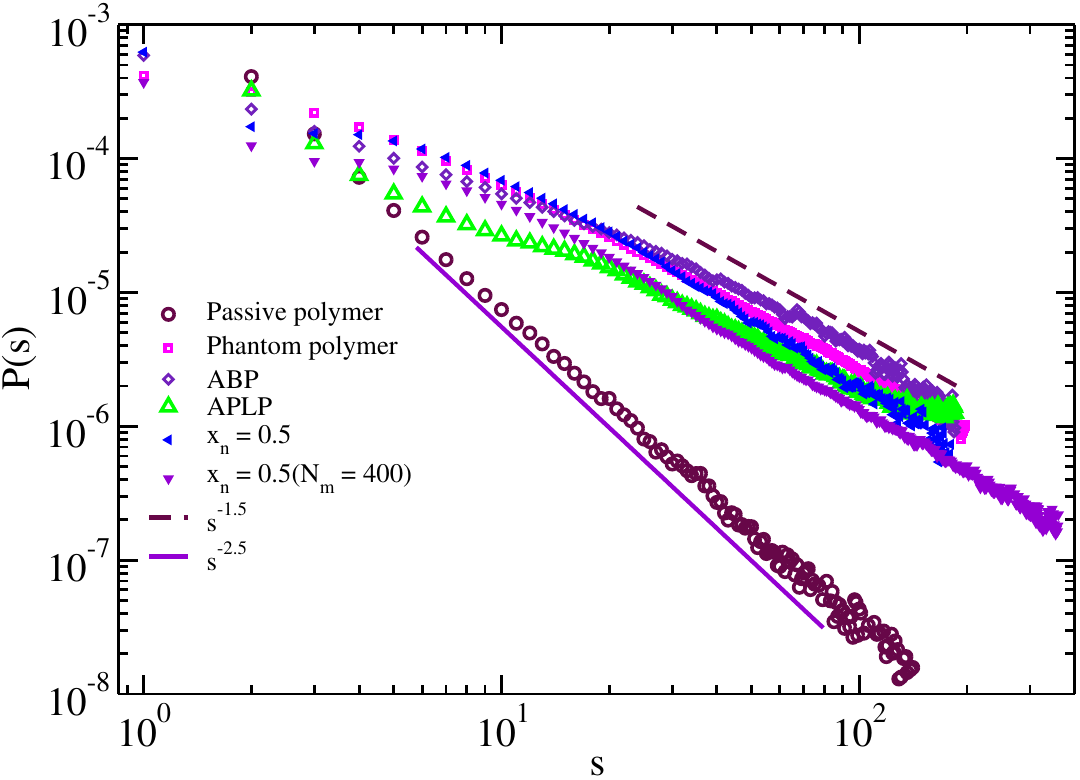}
	\caption{\cblue The normalized contact probability distribution $\mathrm{P(s)}$ of monomers as a function of contour length $s$ shows a transition from self-avoiding to phantom polymer.  Solid and dashed lines indicate power-law behaviors of $s^{-2.5}$  (self-avoiding chain) and $s^{-1.5}$ (phantom chain), respectively at $\mathrm{Pe}=100$. The probability distribution indicates a shift in contact frequency from lower values for the passive polymer to higher values for the active polymers. } 
	\label{Fig:contactmap}
\end{figure}

\subsection{Contact Probability Distribution}
 To understand the underlying physical mechanism that leads to change in the scaling exponent from self-avoiding polymer to ideal polymer, we analyze the contact probability of monomers of active polymer and compare it with that of passive phantom and self-avoiding polymers (real polymers).  
{\cblue The contact probability quantifies the probability of two monomers of a polymer, separated by a distance $s$ along the contour, coming into a spatial proximity of $r_c$. The cutoff distance $r_c = 1.2\sigma$ defines the threshold below which two monomers are considered to be in contact. }

A passive self-avoiding polymer is relatively more stretched than a phantom polymer; therefore, the contact probability of monomers within a polymer is significantly smaller. The estimated contact probability of a real polymer follows a scaling law, $P(s) \sim s^{-2.5}$\cite{natesan2021active}. On the other hand, in a phantom polymer, where excluded volume interactions are absent,  monomers can pass more frequently. This causes the scaling exponent to be smaller, as contact probability is given by $P(s) \sim s^{-d/2}$  where $d$ is the spatial dimension. For $d=3$, the exponent is $ 1.5$ \cite{natesan2021active}. The dashed and solid lines illustrate the power law variation of phantom ($s^{-1.5}$) and self-avoiding polymers ($s^{-2.5}$) in Fig.~\ref{Fig:contactmap}.

Now, we compare the results of the  APLP, ABP, and bidirectional active polymers with those of the two passive polymers.  The APLP exhibits significant compression compared to its passive counterpart in the limit of large activity; therefore, the contact probability is expected to be higher than that of a passive polymer. The results of the APLP compare well with the contact frequency of the phantom polymer $P(s)\sim s^{-1.5}$. However, as the fraction of negatively polarized bonds increases $ (x_n)$, the polymer swells, and its behavior transitions from a compact conformation to a more extended structure.  Even at $ x_n = 0.5$ and in the case of ABP, the polymer remains significantly stretched relative to a passive polymer. Despite extended conformations, the contact probability of bidirectional active polymer exhibits the scaling behavior of a phantom polymer, i.e., $P(s) \sim s^{-1.5}$\cite{natesan2021active}, see Fig.~\ref{Fig:contactmap}.  {\cblue Interestingly, the observed power-law exponent of $P(s)$ is smaller for active polymers, which may substantially decrease in the presence of attractive interactions between different polymer segments and a smaller fraction of active bonds. In such a scenario, our results and model may offer insights into the contact frequency exponents of chromatin fibers, which typically range from 0.75 to 1.5.\cite{lieberman2009comprehensive,chan2024activity,schwarzer2017two,krietenstein2020ultrastructural,hsieh2020resolving}. } 

{\cblue The primary reason for this seemingly non-intuitive behavior for the ABP and bidirectional polymer can be attributed to the activity-induced enhanced diffusivity of monomers.  Although these polymers display distinct structural behaviors, due to their enhanced diffusivity, monomers frequent collision during conformational changes\cite{anand2020conformation,winkler2020physics}.  This leads to a higher contact probability in ABP and bidirectional polymers compared to self-avoiding passive polymers. Consequently, their contact probability distributions are higher compared to those of passive polymers and closely resemble those of an ideal polymer. In summary, the polymer in its stretched state also exhibits ideal chain-like contact probability and structural scaling relations with $\nu \approx 1/2$.(see Fig.\ref{Fig:colormap}-a).  }


 \section{Segmental stretching}
The swelling of the polymer due to bidirectional activity can also be understood in terms of the local stretching of its segments. Polar active polymers exhibit inhomogeneous stretching of different segments and break the symmetry of the polymer~\cite{tejedor2024progressive,panda2025activity}. This local stretching is calculated using the square root of the mean square local distance (R-MSLD), defined as $r_s = \sqrt{\langle \Delta \bm r_n^2(s) \rangle}= \sqrt{\langle ({\bm r}_{s+n}-{\bm r}_{s})^2 \rangle}$. Here, $r_s$ represents the average distance between monomers located at positions $s$ and $s+n$ along the polymer contour, where $n > 0$ and $s$ ranges from $1$ to $N_m-n$.

\begin{figure}[t]
    \centering
    \includegraphics[width=0.475\textwidth]{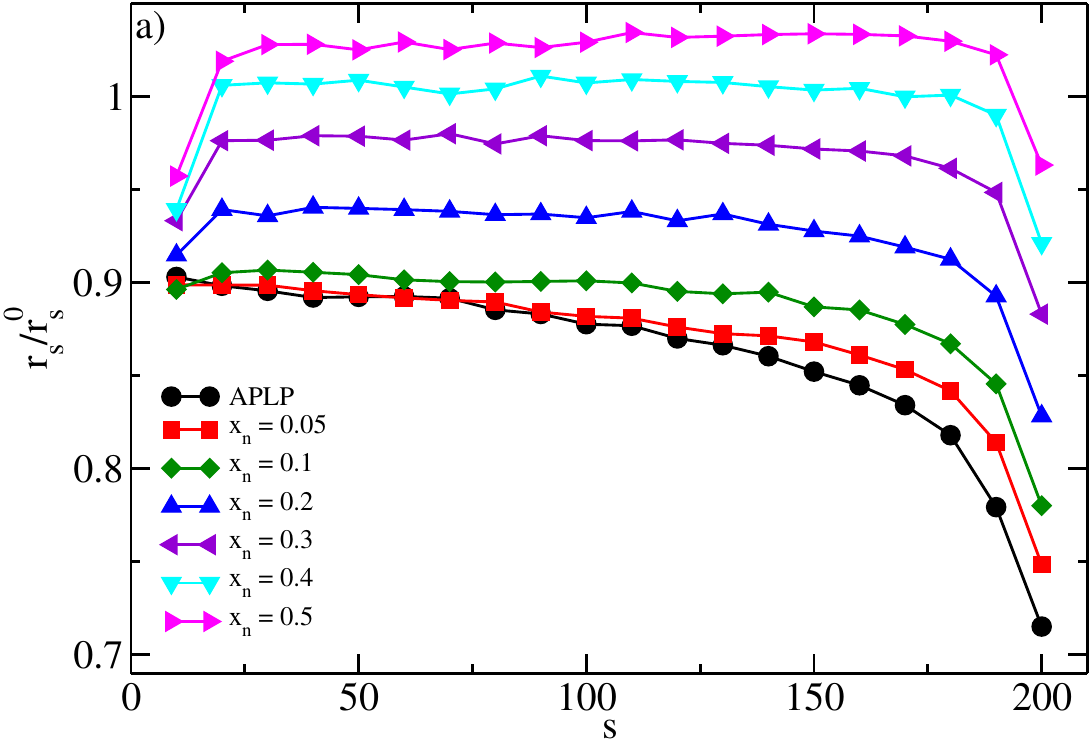}
    \includegraphics[width=0.475\textwidth]{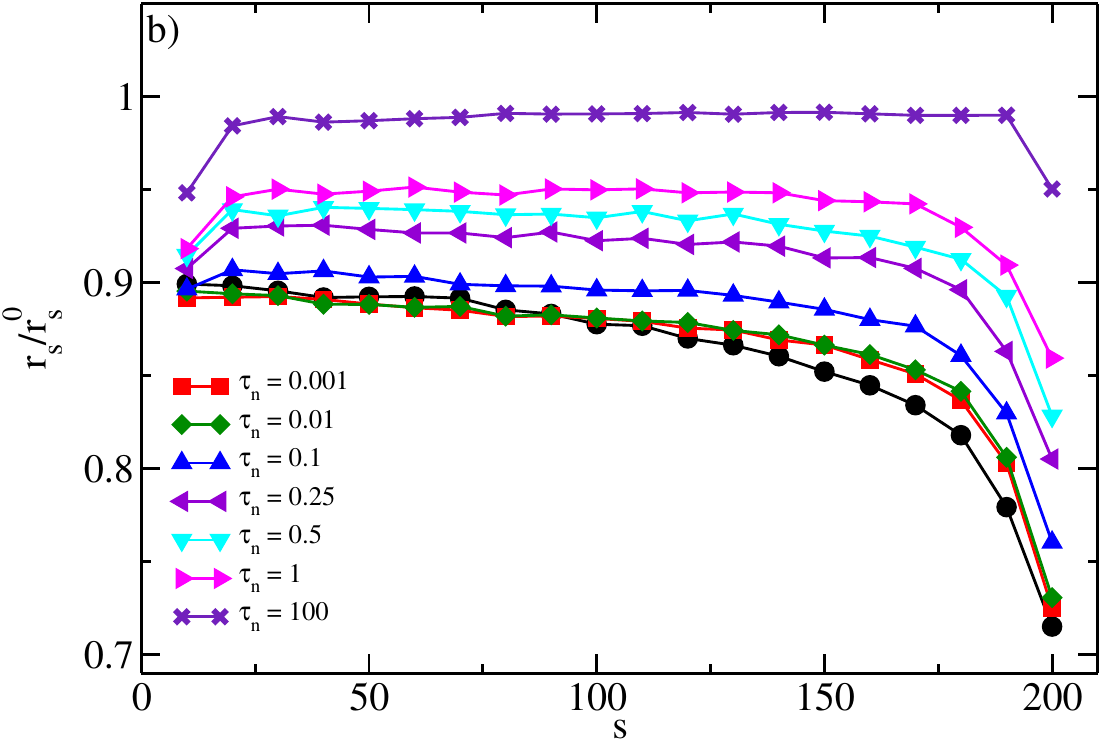}
   \caption{The normalized square root of the mean-square local distance $r_s/r_s^0$ (R-MSLD) of polymer segments of length $n = 9$ as a function of contour length $s$ for $\mathrm{Pe} = 20$ and $N_m=200$. Here, $r_s^0$ is the R-MSLD of the polymer at $\mathrm{Pe}=0$. The normalized R-MSLD $r_s/r_s^0$   for various  $x_n$   at $\tau_n=0.5$ is displayed in sub-figure (a), and sub-figure (b)  presents it for various $\tau_n$ at $x_n=0.2$. The black filled circles represent the results of the APLP model.}
    \label{fig:inhom}
\end{figure}

For a passive polymer ($\mathrm{Pe} = 0$), the polymer is a self-similar structure along its contour, and segmental stretching remains symmetric with respect to its ends. However, in the case of a polar active polymer (APLP), the polymer undergoes anisotropic shrinkage, with a pronounced contraction toward its head end, as shown in Fig.~\ref{fig:inhom}(a) for $x_n = 0$.  

With the increase of negative polarity active bonds ($x_n$), the polymer experiences an elongation force that causes segmental stretching along the contour, see Fig.~\ref{fig:inhom}(a).  This increase in local stretching contributes to the non-monotonic behavior in the polymer's radius of gyration. Additionally, an increase in $x_n$ not only influences the segmental compression but also recovers the symmetry of R-MSLD,  as  Fig.~\ref{fig:inhom}(a) displays curves that are nearly symmetric for $ x_n =0.4$ and  $0.5$.

In addition to the variation in the $x_n$, the polarity reversal time $\tau_n$ can also restore the symmetry in the inhomogeneous compression of the polymer. Figure~\ref{fig:inhom}(b) demonstrates that at $x_n = 0.2$, the polymer's symmetric segmental stretching for large $\tau_n>0.5$. The compression of the polymer near the head decreases as $\tau_n$ increases.   
At small $\tau_n$, the polymer is unable to relax and remains in the state near the APLP.
However, as $\tau_n$ approaches the timescale of monomer diffusion, the polymer segments have enough time to relax, resulting in a reduction in local compression. While in the limit of large $\tau_n$, the polymer exhibits symmetric stretching, resembling the behavior of a passive polymer.

In summary, we have shown here that an increase in the active bonds with negative polarity induces a local extensile force along the contour, promoting local stretching of segments. This stretching grows due to the resulting local extensile force with the increase in $x_n$ and $\tau_n$, which results in swelling of the polymer as presented in Fig.
\ref{Fig:percentage}. This aspect will be further discussed in the bond correlations.

\subsection{Bond-bond correlation}
\begin{figure}[t]
\centering
	\includegraphics[width=0.475\linewidth]{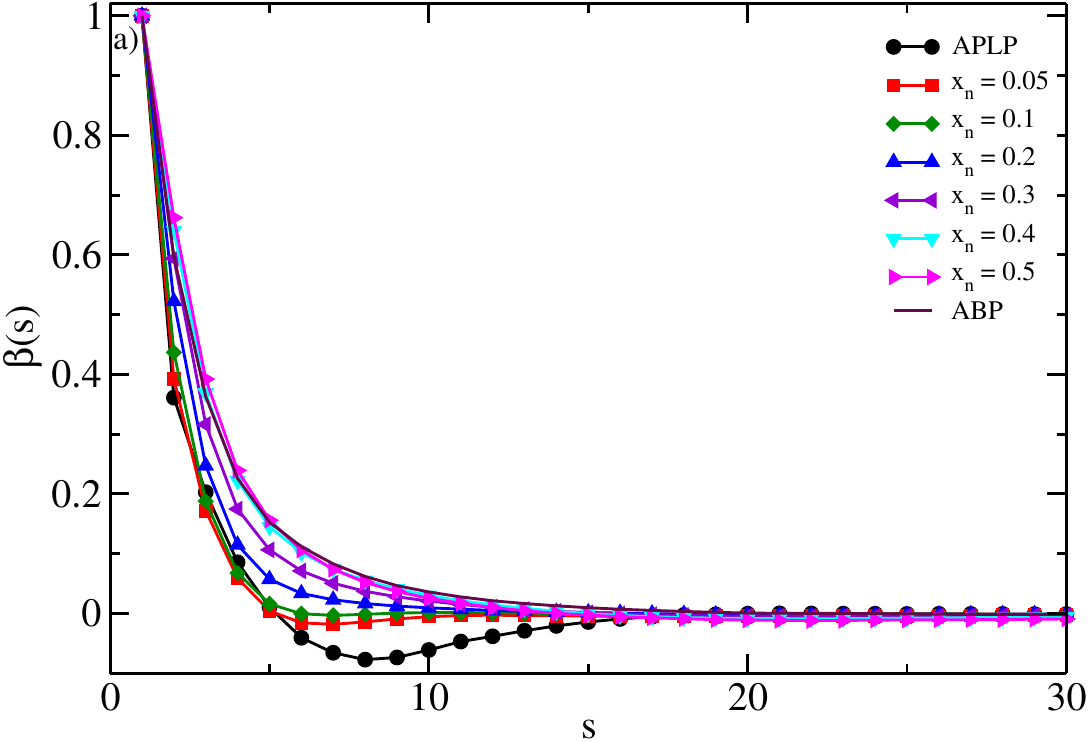}
    \includegraphics[width=0.475\columnwidth]{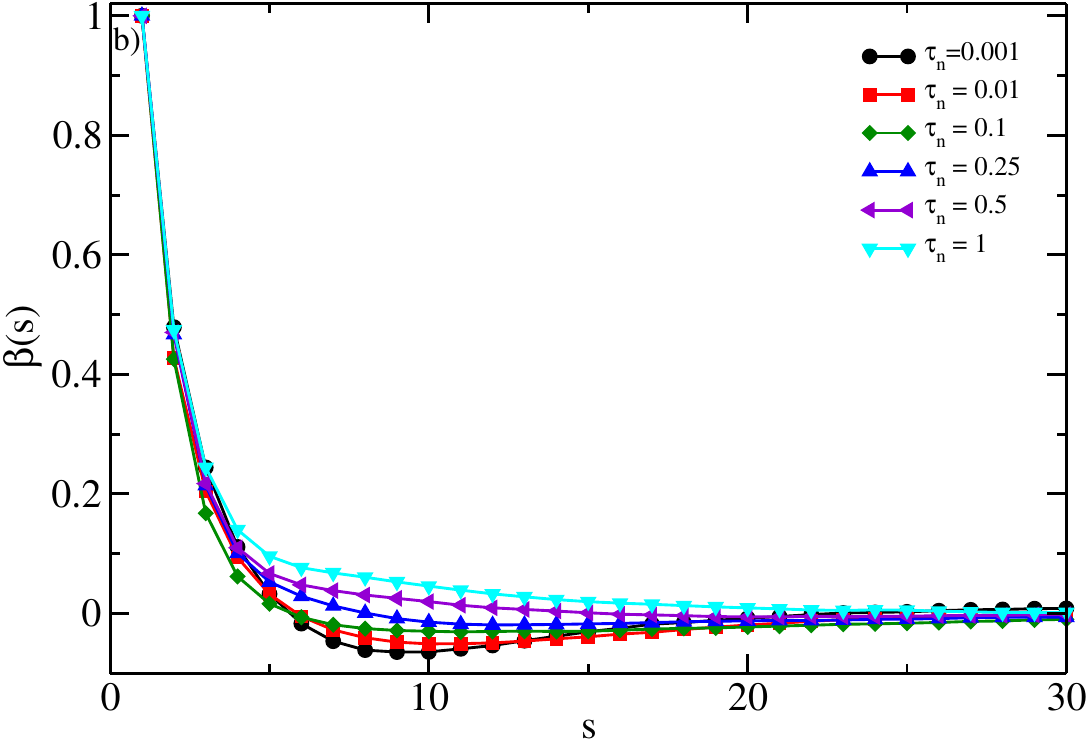}
	\caption{ {\cblue a) The bond-bond correlations  ($\beta(s)= <\mathbf{\hat{t}}_i \cdot \mathbf{\hat {t}}_{j}>$) as a function of contour length  $s=|i-j|\ell_0$  for different values of $x_n$ as indicated in the plot at $\mathrm{Pe} = 50$  and $\tau_n = 0.5$. b)  $\beta(s)$ at $x_n = 0.2$ at various values of $\tau_n$. Figure shows that the bond-bond correlation function exhibits anti-correlation, which disappears for  $x_n>0.1$ and $\tau_n>0.1$. } 
}
	\label{Fig:bond}
\end{figure}

The bond-bond correlation can provide more insights into the compression-to-stretching transition in the polymer's structure. 
To establish this, we estimate the correlation function along the contour of the polymer. The bond-bond correlation function is defined here as,   $\beta(s) = <\mathbf{\hat{t}}_i \cdot \mathbf{\hat {t}}_{j}>$, where $\mathbf{\hat{t}}_i = (\mathbf{r}_{i+1} - \mathbf{r}_{i}) / \lvert \mathbf{r}_{i+1} - \mathbf{r}_{i} \rvert $ represents  the $i^{\mathrm{th}}$  bond vector of unit length, and  $s=|i-j|\ell_0$ contour length.  
  The bond correlation function of a passive flexible polymer quickly decays to zero $\beta(s) \to 0$ with spatial distance ($s$). In contrast, for a tangential active polar polymer, the correlation function $\beta(s)$ reaches zero and further acquires a negative correlation for more than several bond lengths\cite{fazelzadeh2023effects,foglino2019non,vatin2024conformation}. This indicates a local back-folding or compression of the chain under compressive force. The anti-correlation along the chain for smaller $s$ diminishes with the increase in $x_n$.  At $x_n=0.5$,  the correlation function closely resembles that of a correlation of the  ABP polymer model as illustrated in Fig.~\ref{Fig:bond}-(a). 
{\cblue Similarly, Fig.~\ref{Fig:bond}-(b) illustrates the bond correlation for $x_n = 0.2$ at various polarity reversal rates $\tau_n$. At lower reversal rates, the polymer exhibits local back-folding. However, as the reversal rate increases, this local compression gradually disappears, leading to swelling of the polymer. This behavior is consistent with the trend discussed earlier in Fig.~\ref{Fig:percentage}-(b). }
  
  The results of the bond-bond correlation can be used here for the understanding of local stretching. {\cblue In the unidirectional case ($x_n = 0$), all active forces act in the same direction along the polymer backbone, causing local back-folding that leads to overall compression, as discussed in Fig.~\ref{fig:inhom}. However, when bidirectionality is introduced, the active forces no longer align uniformly. Instead, some segments experience forces directed from head to tail, while others experience the opposite, from tail to head. This generates opposing forces along the chain, leading to a buildup of tension, similar to extensile force that effectively pulls different segments in opposite directions along the backbone. Consequently,  local compression turns to the stretching of the polymer at the limit of large $x_n$ and $\tau_n$.  In summary,  the bidirectional active force suppresses anti-correlations and induces local swelling of the chain due to effective extensile forces. This is evidenced by the slower decay of  $\beta(s)$ of the active polymer than that of a passive polymer, see Fig.~\ref{Fig:bond}.  }
  
  

\subsection{Relaxation time} 
The relaxation behavior can be calculated with the end-to-end distance correlation function  $C_e(t)$ that provides a measure of the longest relaxation time of the active polymer\cite{doi1988theory,rubinstein2003polymer, eisenstecken2016conformational,anand2020conformation,weeks1971role,ghosh2014dynamics}.  For $x_n\approx 0$, the correlation function decays and, more importantly, deviates from the standard exponential relaxation in a long time limit, as reported in \cite{tejedor2024progressive}. However, the exponential decay emerges for an equal fraction of positive and negative polarity  $x_n\approx 0.5$, as illustrated in Fig. S1 of the supporting material.\cite{supp} We define the relaxation time $\tau_r$  at which the auto-correlation function has diminished to $\frac{1}{e}$  of its value at $t=0$.  The obtained relaxation time obeys a power-law relation $\tau_r \sim \mathrm{Pe}^{-1}$ in the limit of a large P\'eclet number for the APLP model\cite{bianco2018globulelike,fazelzadeh2023effects,philipps2022tangentially}.  


\begin{figure}[t]
    \centering
    \includegraphics[width=0.475\columnwidth]{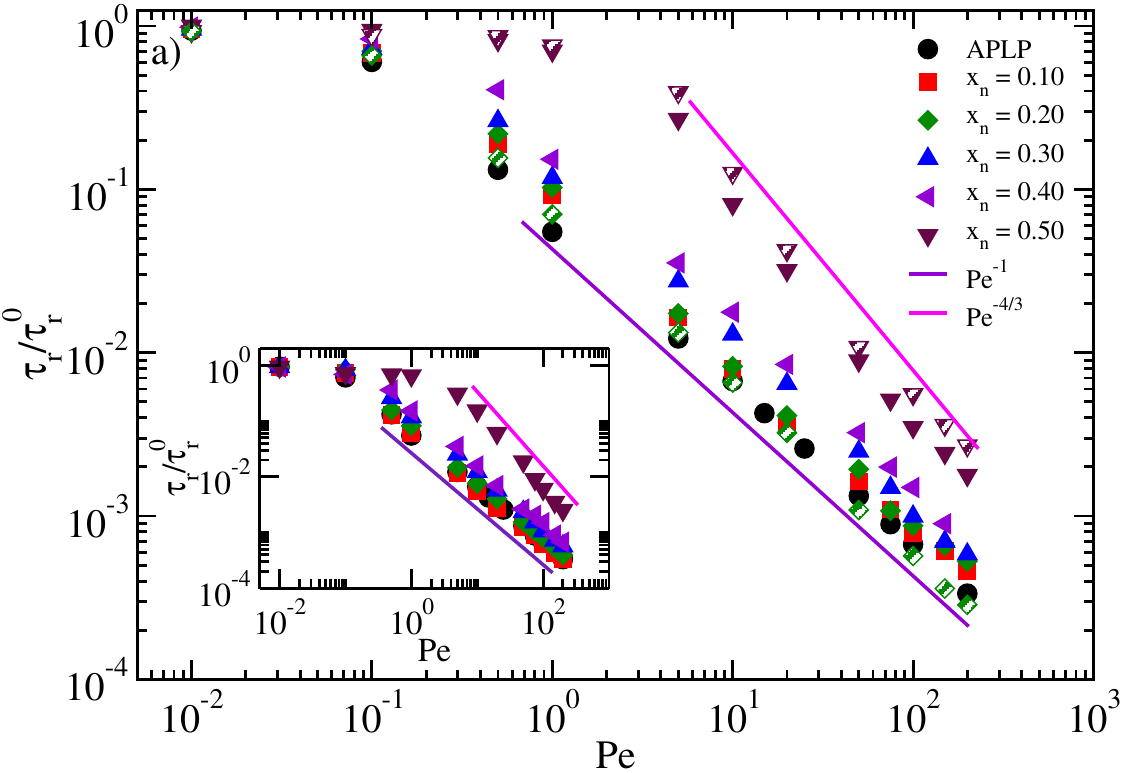}
     \includegraphics[width=0.475\columnwidth]{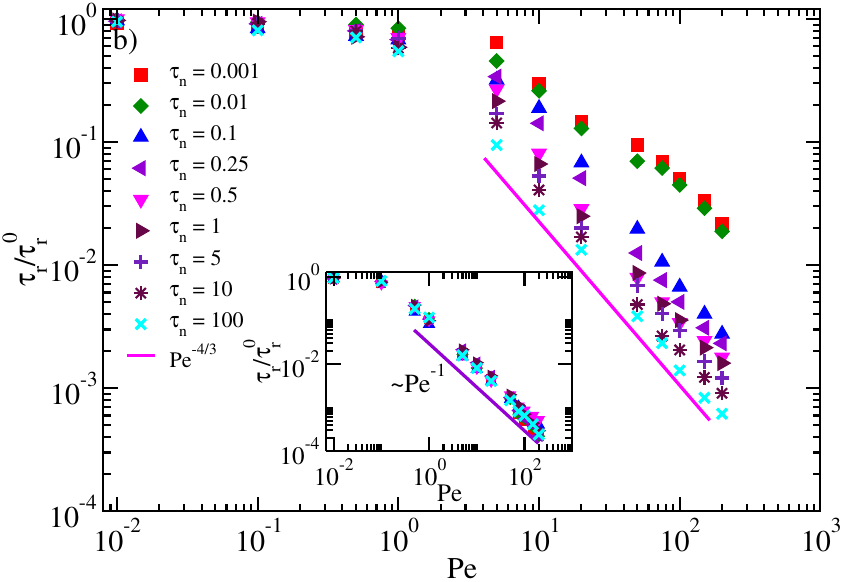}
    \caption{a) The normalized relaxation behavior ($\tau_r/\tau_r^0$) of the active polymer as a function of $\mathrm{Pe}$ on the variation of the polarity of active force ($x_n$) as displayed in the plot for $\tau_n=0.5$ and $N_m=200$ (filled symbols) and $400$ (shaded symbols). The inset depicts scenarios with a relatively small reversal time ($\tau_n = 0.1$). The solid lines guide to the power law behaviors of the relaxation times $\tau_r/\tau_r^0 \sim \mathrm{Pe}^{-1}$ and $\tau_r/\tau_r^0 \sim \mathrm{Pe}^{-4/3}$.   b) The normalized relaxation time  $\tau_r/\tau_r^0$  a function of the P\'elcet number $\mathrm{Pe}$ for various polarity reversal rates.  The main plot corresponds to the equal fraction of bidirectional active force $x_n=0.5$ while the inset represents $x_n=0.2$. Solid lines display power-law behaviors of the relaxation time, as mentioned in the plot.   }
    \label{Fig:relaxatation}
\end{figure}
  
  Figure~\ref{Fig:relaxatation}-a illustrates the significance of the bidirectional active force on the relaxation behavior at a given rate $\tau_n=0.5$ with $\mathrm{Pe}$ and $x_n$. Increase in $x_n$ from zero to $x_n=0.5$,  at equal fraction of the positively and negatively oriented active bonds, a strong deviation from the APLP behavior emerges, leading to a new scaling relation $\tau_r \sim \mathrm{Pe}^{-4/3}$.  The obtained exponent of $4/3$ in this regime is reminiscent of the relaxation behavior observed for the ABP polymer\cite{das2021coil,anand2020conformation,martin2019active}.
  This distinctive shift from exponent $1$ to $4/3$  is also depicted in the inset of Fig.~\ref{Fig:relaxatation}-a for a different reversal time $\tau_n = 0.1$.

  We have also analyzed the relaxation behavior by varying  $\tau_n$ while keeping  $x_n$ fixed.  For  $x_n=0.5$,  a significant deviation on the variation in $\tau_r$,  is observed compared to the APLP case, as shown in Fig.~\ref{Fig:relaxatation}-b. The normalized relaxation time of the polymer shows a significant decrease with increasing $\tau_n$ at large P\'eclet numbers. A power-law regime,  $\tau_r \sim \mathrm{Pe}^{-4/3}$ with an exponent $4/3$ is achieved in the limit of $\tau_n>0.1$\cite{winkler2020physics,martin2019active,anand2020conformation,eisenstecken2016conformational}. Notably at $x_n=0.2$, the relaxation time is nearly independent of  reversal rate $\tau_n$.  In the limit of $\mathrm{Pe}>1$,  $\tau_r$ decreases with a  power-law $\tau_r  \sim \mathrm{Pe}^{-1}$,  strikingly  for all $\tau_n$\cite{bianco2018globulelike,fazelzadeh2023effects,panda2025activity}. Despite substantial differences in their conformational properties, the relaxation behavior remains analogous to that of the APLP model for all $\tau_n$ at $x_n=0.2$, as illustrated in Fig.~\ref{Fig:percentage}-b.  

  In summary, we have shown that the relaxation time of the bidirectional active polymer at $x_n=0.5$ acquires the relaxation behavior of the ABP polymer for the reversal time $\tau_n>0.1$. 

\subsection{Effective Diffusivity}
The bidirectionality of active force influences the transport coefficients of the polymer. To highlight the differences between the unidirectional polar active force and the bidirectional tangential active force, we quantify the dynamical quantities of the polymer. For this we use the mean-squared displacement (MSD) of the polymer's center of mass (CM),  $\langle \Delta r_{cm}^2(t) \rangle = \langle [{\bm r}_{cm}(t) - {\bm r}_{cm}(0)]^2 \rangle$, with the angular brackets ($<..>$) refer time and ensemble averages.  The MSD of the active polymer can be categorized into two distinct time regimes. The short-time limit is called the ballistic or super-diffusive regime, described by $\langle \Delta r_{CM}^2(t) \rangle \sim  t^2$.
In the long-time limit, the MSD of the active polymer attains the diffusive regime, which can be expressed as $\langle \Delta {\bm r}_{cm}^2(t) \rangle \approx 6D_p t$, with $D_p$ denoting the effective diffusivity of the active polymer.

\begin{figure}[t]
    \centering
    \includegraphics[width=0.475\columnwidth]{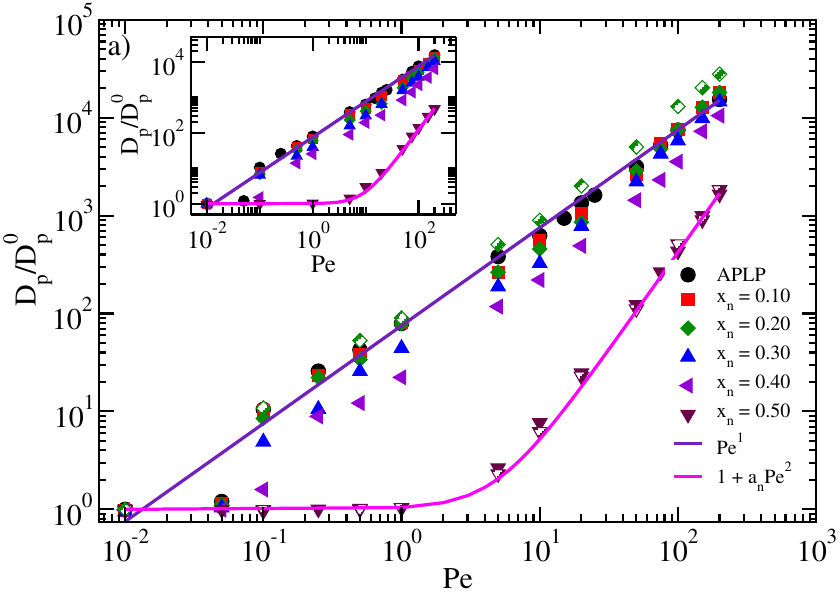}
     \includegraphics[width=0.475\columnwidth]{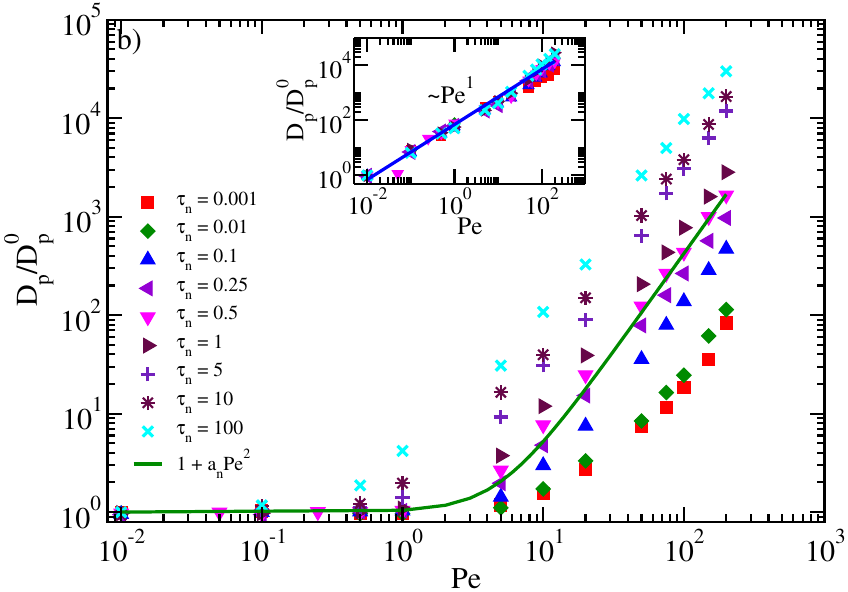}
    \caption{ a) The normalized effective diffusion coefficient ($D_p/D_p^0$) of the active polymer as a function of  $\mathrm{Pe}$ on the variation in the polarity of active force ($x_n$) as displayed in the plot at $\tau_n=0.5$ and $N_m=200$ (filled symbols) and $400$ (shaded symbols). The inset depicts results at reversal time $\tau_n = 0.1$. Solid lines guide the power-law behaviors of the effective diffusivity $D_p/D_p^0 \sim \mathrm{Pe}$ and $D_p/D_p^0 \sim \mathrm{Pe}^{2}$. b) 
     The normalized effective diffusivity $D_p/D_p^0$  as a function of $\mathrm{Pe}$ for various $\tau_n$, as cited in the plot.  The main plot corresponds to $x_n=0.5$, the inset represents $x_n=0.2$, and solid lines display the power-law behaviors of the effective diffusivity.}
     \label{Fig:per_diffusion}
\end{figure} 

We compute the effective diffusivity  ($D_p$) of the active polymer in the diffusive regime. The estimated values of  normalized effective diffusivity $D_p/D_p^0$ for the bidirectional active polymer are presented in 
Fig.~\ref{Fig:per_diffusion} with the variation in $x_n$ and P\'eclet number ($\mathrm{Pe}$) at  $\tau_n=0.5$.  Here, $D_p^0$ is the diffusivity of the passive polymer at $\mathrm{Pe}=0$.  For the APLP case, the effective diffusion coefficient grows linearly  $D_p/D_p^0 \approx (1+ a_p \mathrm{Pe})$  with increasing $\mathrm{Pe}$\cite{philipps2022tangentially,bianco2018globulelike,fazelzadeh2023effects,anand2018structure}, with $a_p$ is a some constant.   The linear increase of the effective diffusivity remains consistent for all $\tau_n$, even though a large fraction of bonds have negative polarity up to $x_n\leq 0.4$. However, a minor deviation in the values of $D_p$ with $x_n$ can be observed in Fig.~\ref{Fig:per_diffusion}.  Interestingly, at $x_n=0.5$, a sharp drop in the effective diffusivity appears and the algebraic form of the effective diffusivity changes from linear to quadratic growth, displaying the following relation  $D_p =D_p^0(1 + a_n \mathrm{Pe}^2)$\cite{bechinger2016active,martin2019active,anand2020conformation}. This behavior is once again one of the main characteristics of the APB polymer. The various solid lines guide the linear and quadratic behaviors of the effective diffusivity in Fig.~\ref{Fig:per_diffusion}. The inset of Fig.~\ref{Fig:per_diffusion} illustrates the effective diffusivity for the different reversal rate $\tau_{n}= 0.1$, exhibiting similar characteristics to the main plot.

The APLP  to the ABP polymer behavior is obtained on the variation in $x_n$, $\mathrm{Pe}$,  and $\tau_n$. The transition from the APLP to ABP polymer is nearly at $x_n=0.5$ for the reversal rate $\tau_n>0.1$. At the same time, when $x_n<0.4$, the bidirectional active polymer retains characteristic features of the APLP model. 

To see the influence of the reversal rate $\tau_n$ on the effective diffusivity $D_p$, we vary $\tau_n$ while keeping $x_n$ fixed.  Fig.~\ref{Fig:per_diffusion}-b shows the effective diffusivity for various $\tau_n$ at $x_n=0.5$ (main plot) and $0.2$ (inset). This plot illustrates that  $D_p/D_p^0$ is linearly increasing for the presented range of $\tau_n$ with $\mathrm{Pe}$. At $x_n=0.5$, the effective diffusivity shows a quadratic increase with $\mathrm{Pe}$ for $\tau_n>0.1$. For the large reversal time,  we obtain the ABP behavior near an equal fraction of positive and negative polar active bonds ($x_n\approx 0.5$).

In summary, at $x_n=0.5$ and $\tau_n\geq 0.1$, the effective diffusivity of the polymer increases quadratically with $\mathrm{Pe}$.
The effective diffusivity is also determined from the active force correlation; the obtained values display similar characteristic features, see SI-Fig. S3. This transition can be understood through force correlation, which shows an anti-correlation similar to the bond-bond correlation observed in the APLP model. As expected, this anti-correlation vanishes for the case of bidirectional active polymers, indicating diffusive behavior akin to ABP polymer-like SI-Fig.S3\cite{supp}.

\section{ Stochastic Polarity  Reversal}
So far, we have assumed that polarity reversal events occur at fixed time intervals; however, these events are inherently stochastic. To account for this aspect, we also considered a scenario in which $\tau_n$ and $\tau_p$ are sampled from a Poisson process, with mean event rates $\lambda_n$ and $\lambda_p$, respectively.  Thus,  the fraction of negatively polar active bonds dynamically fluctuates around $x_n$  at each instant based upon the ratio $x_n = \frac{\lambda_n}{\lambda_n + \lambda_p}$.  

\begin{figure}[t]
    \centering
    \includegraphics[width=0.475\textwidth]{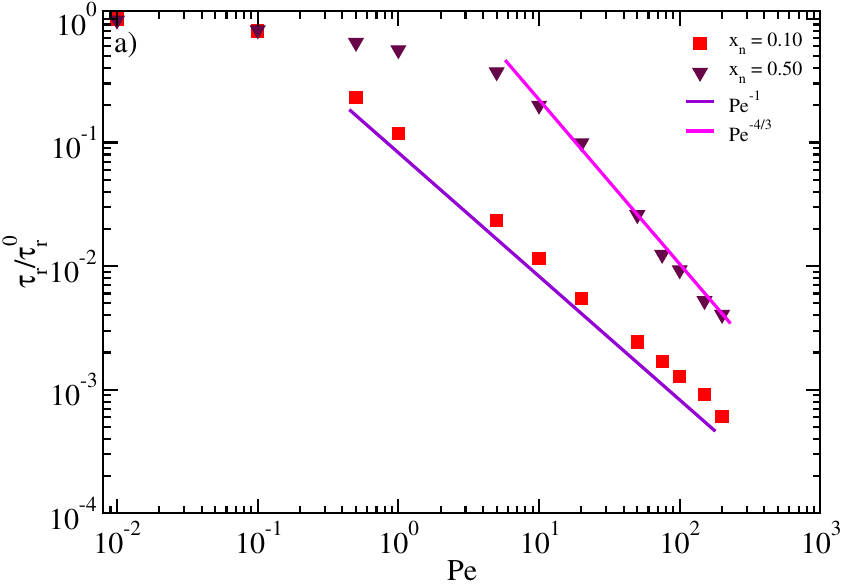}
    \includegraphics[width=0.475\textwidth]{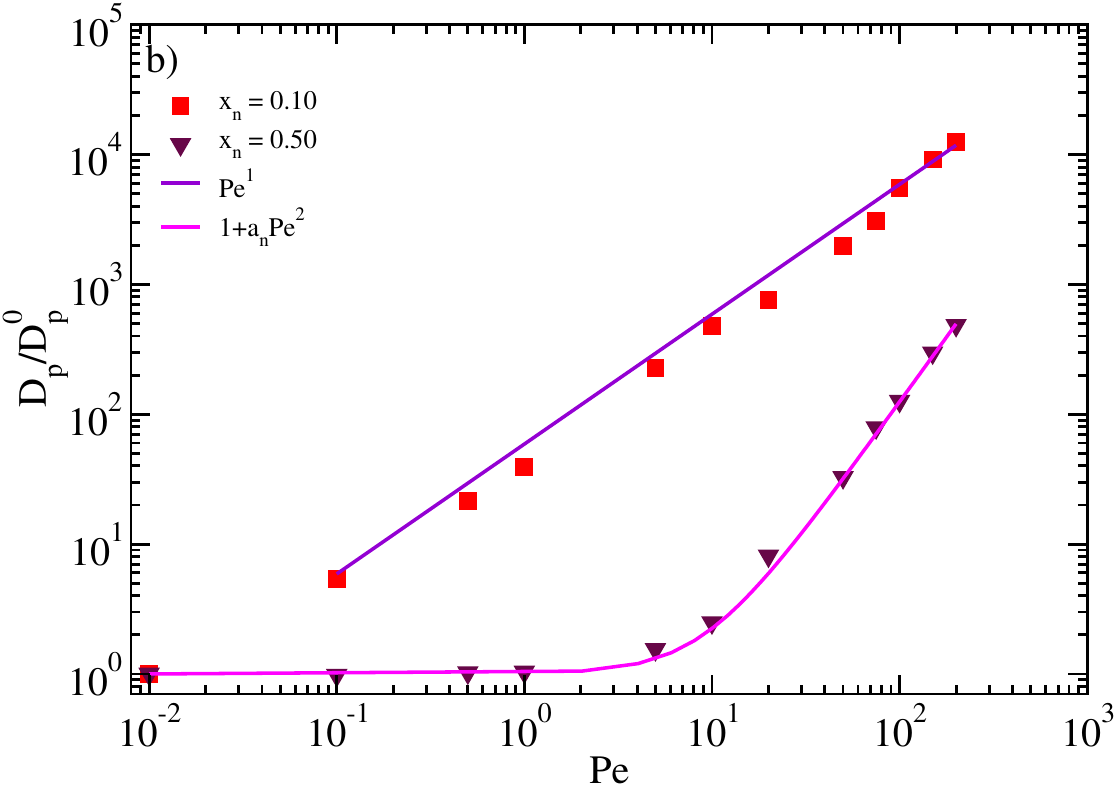}
   \caption{\cblue (a) Figure shows normalized relaxation time $\tau_r / \tau_r^0$ of the active polymer as a function of $\mathrm{Pe}$, at $\tau_n = 0.1$ and  $N_m = 200$. The solid lines illustrate the power-law dependence of the relaxation time, highlighting two distinct power-law regimes: $\tau_r/\tau_r^0 \sim \mathrm{Pe}^{-1}$ and $\tau_r/\tau_r^0 \sim \mathrm{Pe}^{-4/3}$.  
(b) The normalized effective diffusion coefficient  $D_p/D_p^0$ of the active polymer as a function of $\mathrm{Pe}$ for various $x_n$ at $\tau_n = 0.1$ and $N_m = 200$. The solid lines demonstrate the linear and quadratic scaling behaviors of the normalized effective diffusivity.}
    \label{fig:time_diff}
\end{figure}

 Under this framework, the structural and dynamic properties of the polymer are also investigated and presented here for comparison.  Figures~\ref{fig:time_diff}(a) and (b) illustrate the behavior of the normalized relaxation time $\tau_r/\tau_r^0$  and effective diffusivity $D_p/D_p^0$ of the polymer for two values of $x_n = 0.1$ and $x_n = 0.5$ at a reversal rate $\tau_n = 0.1$.  For $x_n = 0.1$, the relaxation time decreases linearly with activity, following the scaling relation $\tau_r/\tau_r^0 \sim \mathrm{Pe}^{-1}$ similar to the case presented in  Fig.\ref{Fig:relaxatation}-a. Similarly, for $x_n = 0.5$, the relaxation time exhibits a steeper decay as shown in Fig.\ref{Fig:relaxatation}-a, adhering to the power-law behavior $\tau_r/\tau_r^0 \sim \mathrm{Pe}^{-4/3}$.  
 Additionally,  the effective diffusivity increases linearly with activity,  as $D_a/D_p^0 \sim \mathrm{Pe}$, however, for $x_n = 0.1$, and, $x_n = 0.5$, displays quadratic increase on $\mathrm{Pe}$, described by $D_p/D_p^0 = 1 + a_n \mathrm{Pe}^2$, where $a_n \approx 0.5$. These behaviors are similar to the results reported in the previous sections on the effective diffusivity and relaxation time. Thus, in summary, the stochastic polarity reversal rate shows qualitatively similar behavior in various physical quantities as observed for a fixed polarity reversal rate.

\section{Conclusion}
 We have presented the structural and dynamical properties of the bidirectional tangential active polymer and compared its various physical quantities with the APLP and  ABP polymer models.   
 The structural and dynamical transition is governed by the variation of bonds of negative polarity $x_n$ and their reversal time $\tau_n$. In the case of APLP, the compressive active force induces local compression, leading to an ideal polymer-like scaling exponent of $\nu \approx 1/2$ in the limit of large activity, consistent with the previous findings\cite{tejedor2024progressive,bianco2018globulelike}. In contrast, when the bidirectional tangential active force is applied ($x_n > 0$), the compressed polymer substantially swells, particularly at  $x_n=0.5$ in the large activity limit ($\mathrm{Pe}>>1$). This yields a non-monotonic structural behavior, which is further accentuated by polarity reversal time ($\tau_n$). Large reversal time $\tau_n>0.1$ suppresses the compressive behavior of the polymer and enhances the stretching.

Additionally, we have shown that the polymer exhibits purely monotonic swelling in the limit of very large $\tau_n>>1$. Notably, our simulations have also demonstrated that while the reversal rate $\tau_n$ significantly influences the polymer conformation, the dynamical behavior remains essentially unchanged for the asymmetric distribution of the polarity $x_n=0.2$.  Furthermore, we have shown that in the swollen state, the polymer's conformations exhibit the ideal-chain-like scaling with the exponent $\nu\approx 1/2$. This has also been seen in previous theoretical results of the ABP polymer, revealing the scaling exponent $\nu=1/2$ in the stretched states\cite{winkler2020physics,eisenstecken2016conformational,peterson2020statistical}. 
 
The structural transition and insights into the polymer's structure are provided here through bond-bond correlations, inhomogeneous stretching, and monomer contact frequencies. The bond-bond correlations reveal that local compression and folding of the polymer are suppressed in the bidirectional active polymer due to the local effective extensile force. As a result, local stretching and correlations become comparable to those observed in ABP models under bidirectional active forces.   Furthermore, the contact frequencies of monomers in both compressed and stretched conformations closely resemble those of a passive phantom polymer, suggesting that the polymer exhibits the ideal chain-like scaling in these states.


The transition from the APLP to the ABP models is further apprehended from the dynamical and relaxation behaviors. The relaxation time of the polymer decreases linearly even with an increase in the fraction of the active bonds up to  $x_n>0.4$; however, in the regime of nearly equal polarity,  relaxation follows as  $\tau_r  \sim \mathrm{Pe}^{-4/3}$ (ABP polymer model)—similarly, a linear increase of the effective diffusion coefficient changes to a quadratic one with $x_n$. This transition is further comprehended from the active force correlation, which exhibits the anti-correlation feature in its relaxation. This is the consequence of a change from a compressive force to a random active force\cite{winkler2020physics}.


Our model offers a simple alternative to polymer compaction without attractive interactions between beads, more importantly, that can be reversibly unfolded by tuning the polarity reversal rate or introducing a polarity imbalance.  
A polymer's ability to fold and unfold under active force is essential for diverse biological functions, for example, compaction of the chromatin fiber in the nucleus and unfolding of various proteins.  It is established that chromatin compaction occurs via loop extrusion driven by specific protein complexes. In the loop extrusion, chromatin is pulled bidirectionally for compaction.\cite{misteli2020self}  
Additionally, in eukaryotic cells, cargo transport along microtubule tracks by opposing molecular motors is often bidirectional, generating active stresses that significantly influence the mechanical and transport properties of cytoskeleton filaments.  Although our study focuses on a generic model of bidirectional active polymers with all bonds active, it can be extended to more specific systems—such as chromatin compaction by protein complexes or RNA polymerase-driven motion along DNA, and mechanical and transport properties of cytoskeleton filaments, where only a subset of bonds is active in a bidirectional and stochastic manner.

The proposed theoretical model of a bidirectional active polymer poses challenges but remains feasible for a synthetic polymer. Experimental studies have revealed that the direction of activity of individual colloids can be reversed by tuning parameters such as wavelength, temperature, laser intensity, or magnetic field\cite{vutukuri2017rational,gomez2017tuning,ji2019thermoresponsive,garcia2018magnetically}. Polymer-like chains formed from emulsion droplet-based active particles, chemically powered Janus colloids, or self-assembled hybrid nanorod-spheres have already been realized, and they demonstrate active polymer-like behavior.\cite{kumar2024emergent,biswas2017linking,garcia2018magnetically}.
Therefore, regulating the self-assembly of these colloidal suspensions and switching the direction of activity through suitable physical parameters can open up promising avenues for realizing bidirectional active polymers\cite{vutukuri2017rational,kumar2024emergent,mcmullen2018freely,biswas2017linking}. 

At the macroscopic scale, a polymer chain made of granular mini-motors represents another promising experimental platform for mimicking the behavior of bidirectional active polymers.  The study of such chains, whose beads exhibit circular chiral motion, has demonstrated spontaneous folding–unfolding transitions in two dimensions \cite{caprini2024spontaneous}. Designing active granular beads with stochastic directional switching based on front- or rear-bead sensing could offer new opportunities to explore active polymer dynamics at the macroscopic scale.

\begin{suppinfo}
This contains the end-to-end correlation function, bond-bond correlation function, active force correlation, and a flow chart of the simulation model, illustrating the implementation of polarity reversal within the framework of overdamped Langevin dynamics.\cite{supp,panda_2025_17011905}

\end{suppinfo}

\begin{acknowledgement}

AP and SPS acknowledge financial support from the DST-SERB Grant No. CRG/2020/000661, IISER Bhopal, and computational time at IISER Bhopal and the Param Himalaya NSM facility.

\end{acknowledgement}

\providecommand{\latin}[1]{#1}
\makeatletter
\providecommand{\doi}
  {\begingroup\let\do\@makeother\dospecials
  \catcode`\{=1 \catcode`\}=2 \doi@aux}
\providecommand{\doi@aux}[1]{\endgroup\texttt{#1}}
\makeatother
\providecommand*\mcitethebibliography{\thebibliography}
\csname @ifundefined\endcsname{endmcitethebibliography}
  {\let\endmcitethebibliography\endthebibliography}{}

\end{document}



\title{Supplementary Material: Folding-unfolding transition  of  active polymer on the reconfiguration of bidirectional tangential active force}

\author{Arindam Panda}
\email{arindam19@iiserb.ac.in}
\affiliation{Department of Physics,\\ Indian Institute of Science Education and Research, \\Bhopal 462 066, Madhya Pradesh, India}
\author{Sunil P. Singh} 
\email{spsingh@iiserb.ac.in}

\affiliation{Department of Physics,\\ Indian Institute of Science Education and Research, \\Bhopal 462 066, Madhya Pradesh, India}



\date{\today}

\maketitle

\begin{figure}[t]
\centering
	\includegraphics[width=0.48\columnwidth]{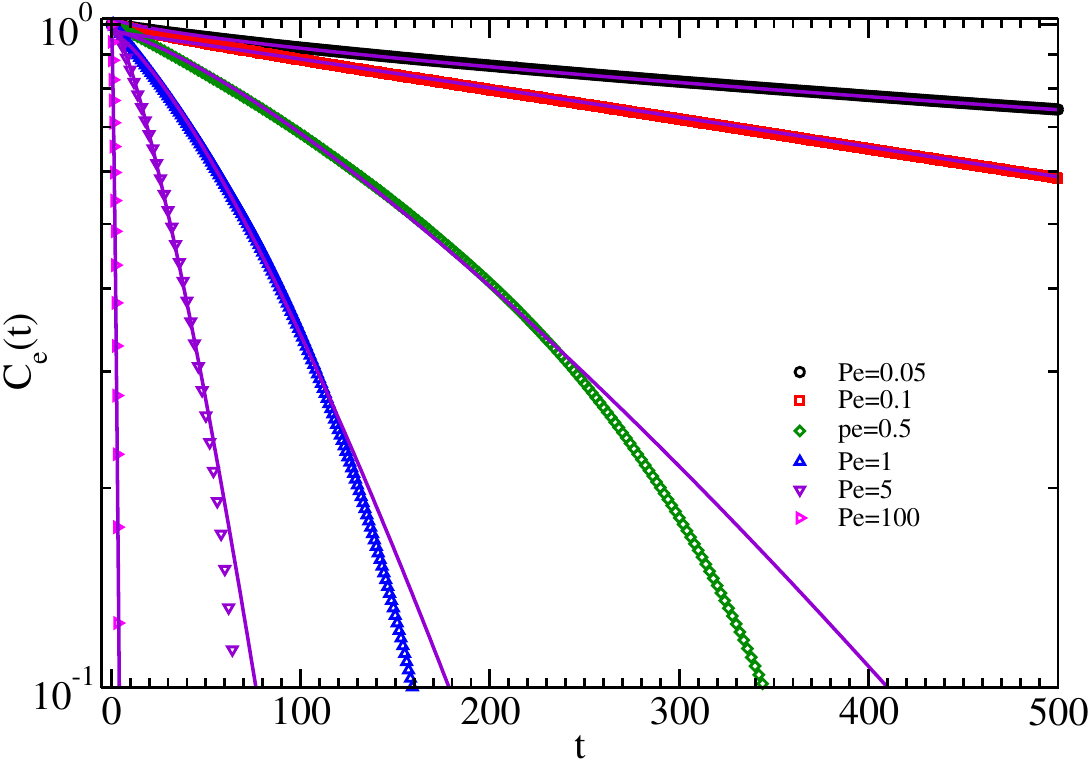}
    \includegraphics[width=0.48\columnwidth]{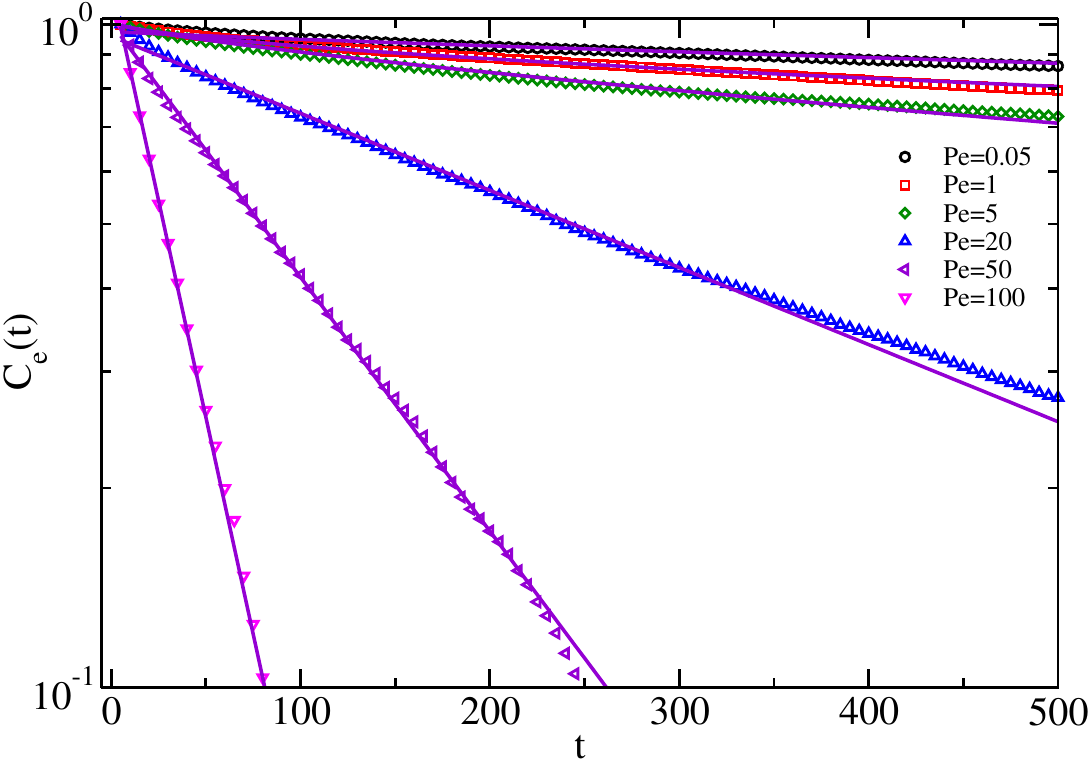}
	\caption{The figure displays the autocorrelation function of the end-to-end vector for the APLP case (a) and for bidirectional active polymer $x_n=0.5$ (b) at $N_m=200$ for various $\mathrm{Pe}$. The solid lines display the exponential fit to the correlation function given as $C_e(t)\approx \exp(-t/\tau_r)$.  }
	\label{Fig:autocorr}
\end{figure}
\section{Relaxation of polymer}
The longest relaxation time $\tau_r$ of the polymer is determined from the end-to-end vector autocorrelation function expressed as $C_e(t) = \langle {\bm R}_e(t) \cdot {\bm R}_e(0) \rangle$. Figure~\ref{Fig:autocorr}(a) illustrates the autocorrelation function of the active polar linear polymer(APLP). The results indicate that the autocorrelation function deviates from exponential decay even before it has reached to $  e^{-1}$ value\cite{tejedor2024progressive,fazelzadeh2023effects}.
 
In contrast, Fig.~\ref{Fig:autocorr}(b) presents the relaxation behavior of a bidirectional active polymer at $x_n = 0.5$, a regime where the polymer effectively mimics  ABP behavior. Here, the autocorrelation function decays in a simple exponential manner,  up to $C_e(\tau_r) = e^{-1}$ and beyond too. Given the distinct patterns of the $C_e(t)$ in these two scenarios, the relaxation time $\tau_r$ is defined as the time where the autocorrelation function has decayed to $C_e(\tau_r) = e^{-1}$. This approach provides a unified framework to compare relaxation times across different active polymer systems despite the distinct nature of their autocorrelation functions.

 \section{Bond-bond correlation}
\begin{figure}[t]
\centering
	\includegraphics[width=0.48\columnwidth]{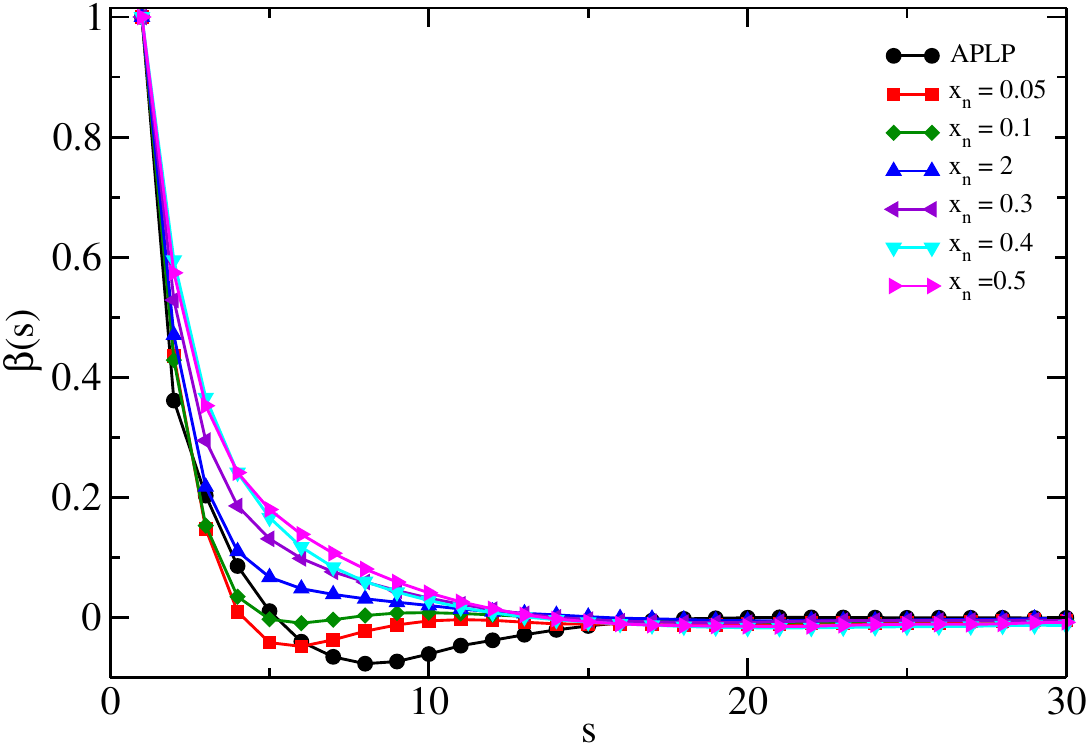}
    \includegraphics[width=0.48\columnwidth]{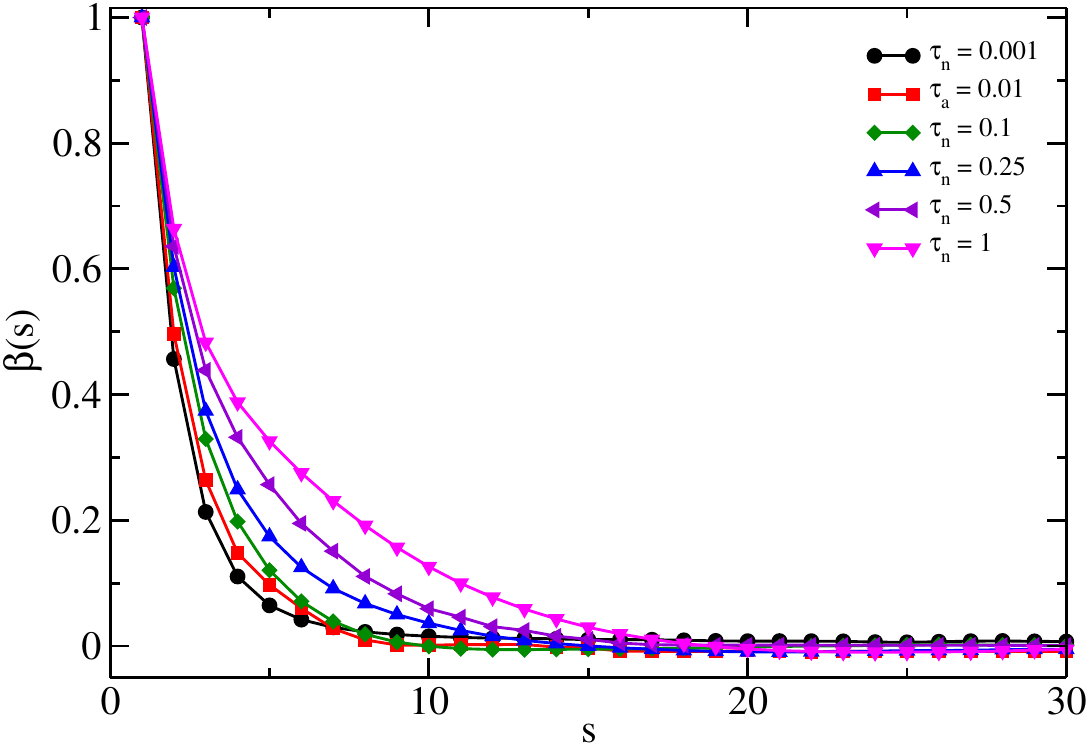}
	\caption{ 
The bond-bond correlation function $\beta(s)$ is plotted as a function of contour length $s$ for (a) various values of $x_n$ at fixed $\tau_n = 0.1$, and (b) different values of $\tau_n$ at fixed $x_n = 0.5$, at activity $\mathrm{Pe} = 50$.}
	\label{Fig:bond}
\end{figure}

 We have included additional plots of the bond-bond correlation function $\beta(s)$ at $\tau_n = 0.1$ for varying values of $x_n$, as well as for a symmetric distribution of active bonds ($x_n = 0.5$) across different $\tau_n$ values. At $\tau_n = 0.1$, $\beta(s)$ shows clear anti-correlation for asymmetric cases, which vanishes when the active bond distribution is symmetric ($x_n = 0.5$). As expected, increasing $\tau_n$ leads to a slower decay of bond-bond correlations for the symmetric case ($x_n = 0.5$), as shown in Fig.~\ref{Fig:bond}. Notably, no anti-correlation is observed in the bond correlations at $x_n = 0.5$.

 \section{Active force autocorrelation}
\begin{figure}[t]
\centering
	\includegraphics[width=0.48\columnwidth]{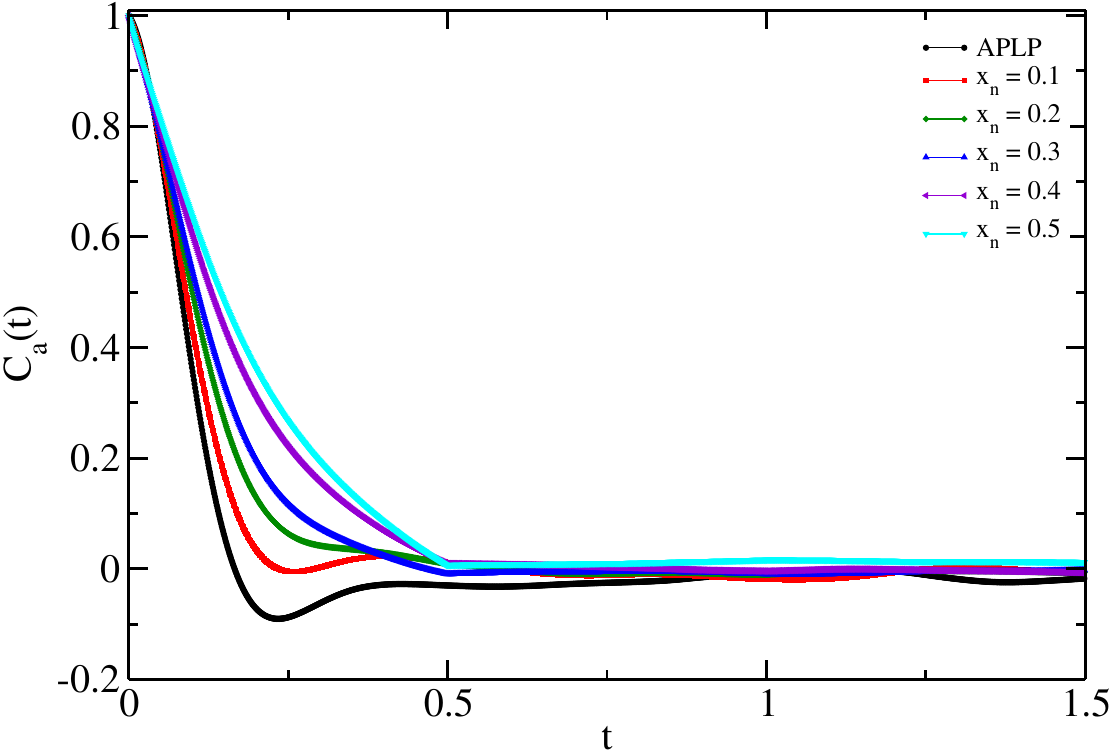}
    \includegraphics[width=0.48\columnwidth]{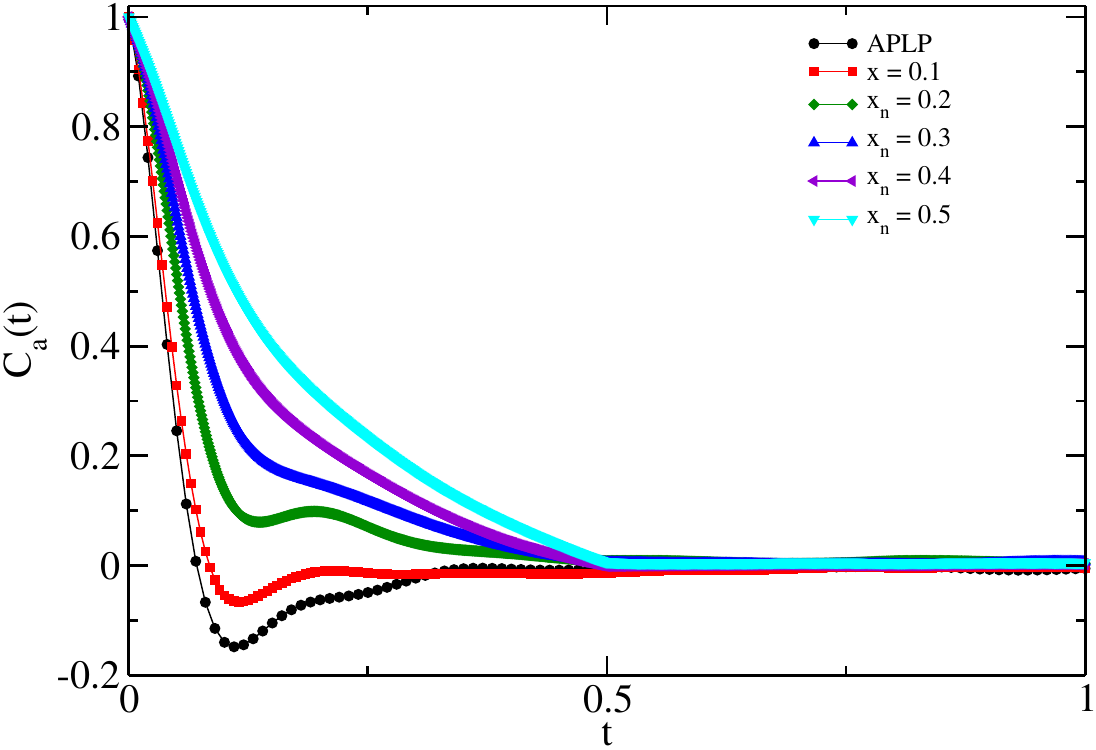}
	\caption{The correlation function of the total active force $C_a(t)$     for varying fractions of negatively oriented  active bonds  at $\tau_n = 0.5$ (a)  $\mathrm{Pe} = 25$ (b) for $\mathrm{Pe} = 50$.}
	\label{Fig:Active_corr}
\end{figure}

 The active polymer's effective diffusivity ($D_p$) can be obtained from the active force autocorrelation function.  The equation of motion of the center of mass of the polymer can be rewritten as, $\zeta {\bm v}_{cm}=\frac{1}{N_m}(f_a\sum_{i=1}^{N_m} p_i \bm{\hat t}_i+ \sum_{i=1}^{N_m} \bm{\Gamma}_i$). This  can be further simplified in the limit of large $\mathrm{Pe}\gg 1$,
  ${\bm v}_{cm} \approx  \frac{f_a}{N_m \zeta }\sum_{i=1}^{N_m-1}  p_i\bm{\hat t}_i\approx {\bm F}^a_{cm}/\zeta$,  here ${\bm F}^a_{cm}$ is the total active force on the polymer's center of mass.
 The total active force on the polymer's center of mass can be separately estimated in the simulations, which can be expressed as  $ {\bm F}^a_{cm} \approx \frac{f_a}{N_m }\sum_{i=1}^{N_m-1}  p_i\bm{\hat t}_i$.  The correlation is estimated  from the expression $C_a(t)=<{\bm F^a}_{cm} (t)\cdot {\bm F^a}_{cm}(0)> $.   The force correlation  for various $\mathrm{Pe}$, $x_n$, and $\tau_n$ is computed. The function $<{\bm F^a}_{cm} (t)\cdot {\bm F^a}_{cm}(0)>$ of the APLP displays anti-correlation in the intermediate time regimes before approaching to zero in the limit of large time\cite{vatin2024conformation}. This negative correlation fades away for the bidirectional active polymer. At $x_n=0.5$, the active force correlation falls sharply, displaying the features near the ABP model; see Fig.~\ref {Fig:Active_corr}. The appearance of the negative correlation is due to the negative curvature along the polymer, as seen in the bond-bond correlation function in Fig.6 (main manuscript). These negative correlations appear to be intrinsic features of the APLP, as polymer does have progressive stretching \cite{tejedor2024progressive,panda2025activity}. 

\begin{figure}[h!]
\centering
	\includegraphics[width=0.5\linewidth]{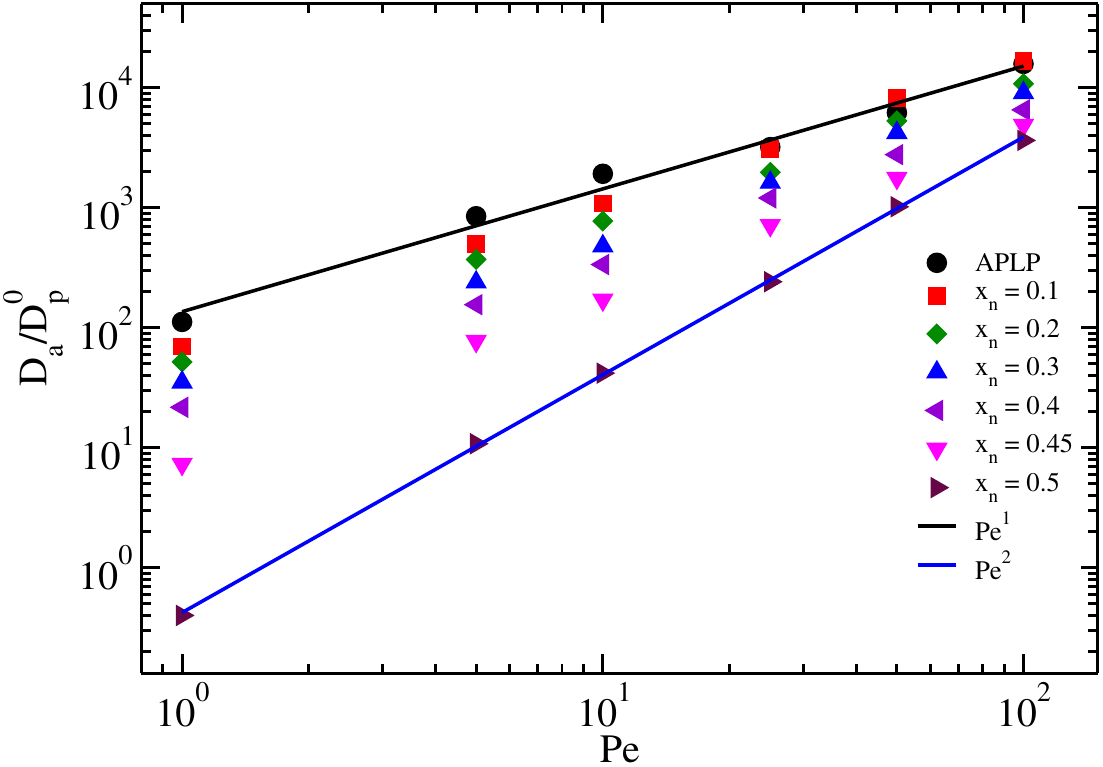}
	\caption{The effective diffusivity $D_a/D_p^0$ of the active polymer obtained from the force correlation as a function of  P\'eclet number for various $x_n$. The solid lines display the linear and quadratic behaviors of the effective diffusivity.  }
	\label{Fig:Active_diff}
\end{figure}

We estimate the effective diffusivity from the correlation simply by integrating $D_a\approx 2 \int_0^{\infty} <{\bm F^a}_{cm} (t)\cdot {\bm F^a}_{cm}(0)>dt $.  The obtained normalized effective diffusivity $D_a/D_p^0$ is presented in Fig.~\ref{Fig:Active_diff} as a function of $\mathrm{Pe}$
 for various $x_n$ at a given $\tau_n$.  Interestingly, the effective diffusivity obtained from this approach also exhibits the linear increase with $\mathrm{Pe}$ for $x_n<0.4$, and for  
 $x_n\approx0.5$,  once again, effective diffusivity attained, quadratic increase with $\mathrm{Pe}^2$.

\begin{figure}[h!]
    \centering
    \includegraphics[width=0.5\textwidth]{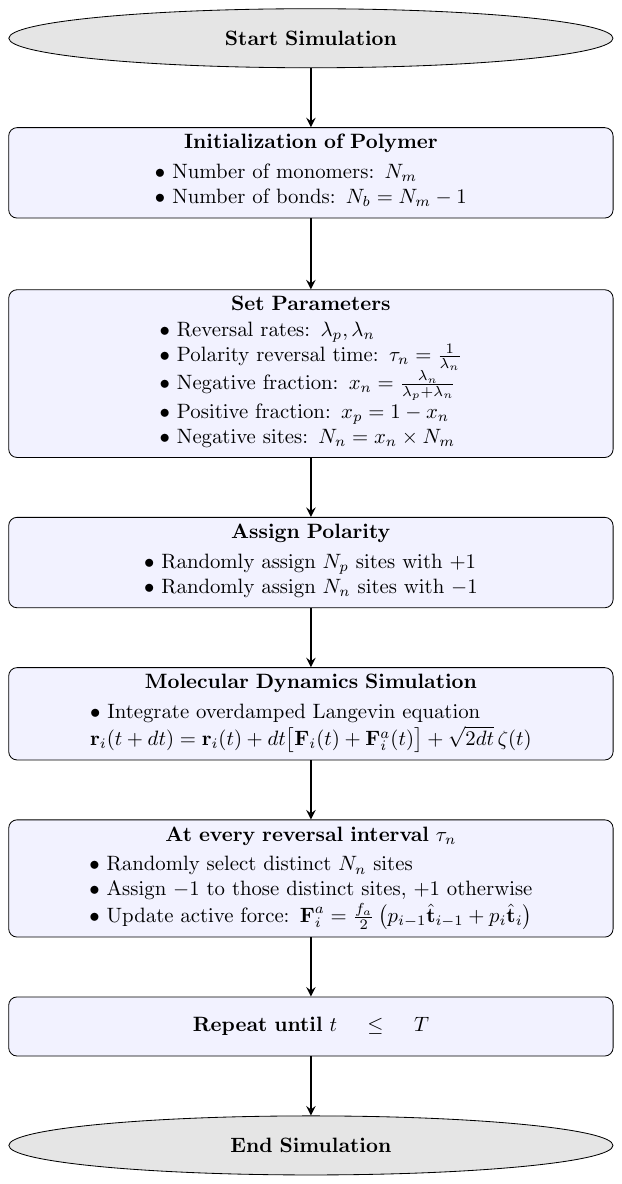}
   \caption{Simulation flowchart displays from the initialization of the simulation parameters, the update of the stochastic change in the orientation of the active force,  and the integration step.  }
    \label{fig:nu}
\end{figure}

\section{Simulation Flowchart}
In this section, we provide a flow chart of the simulation algorithm and a detailed scheme for polarity reversal\cite{panda_2025_17011905}.  The flow chart shows the primary steps and their order. 

{\it Initialization of Polymer:} The simulation begins by initializing a polymer consisting of $N_m$ monomers connected by $N_b = N_m - 1$ bonds.  Several key parameters of the model are given as follows. 

    \begin{enumerate}
        \item Reversal rates $\lambda_{p}$ and $\lambda_{n}$, which govern the active force polarity reversal dynamics.  
        \item Polarity reversal time, $\tau_n = \frac{1}{\lambda_n}$, denotes the interval at which the polarity of bonds is randomly reversed to negative polarity with the probability of $x_n$.  
        \item Fractions of bonds with negative and positive polarity, given by $x_n = \frac{\lambda_n}{\lambda_p + \lambda_n}$ and $x_p = 1 - x_n$, respectively.  
        \item Number of active sites with negative polarity, $N_n = x_n \times N_m$.  
    \end{enumerate}  

{\it Assign Polarity:} We randomly assign polarity to each bond, $+1$ for $N_p$ number of positive active sites $-1$ to remaining $N_n$ number of negatively active sites.

{\it Molecular Dynamics Simulation:}  The molecular dynamics simulation is then carried out by integrating the overdamped Langevin equation. The discrete form of the  position update is given as,  
  \begin{equation}
    \mathbf{r}_i (t + dt) = \mathbf{r}_i (t) + dt \left[ \mathbf{F}_i (t) + \mathbf{F}_i^a (t) \right] + \sqrt{2\,dt}\zeta(t),
    \label{Eq:update}
  \end{equation}
    where thermal noise $\zeta(t)$ has a zero mean and a variance $1$. The integration is performed over the total simulation time $T$ with an integration time step $dt$.  \\

{\it Stochastic Polarity Reversal:} In this part, we discuss how the polarity reversal is implemented at every polarity reversal interval time $\tau_n$. 
    \begin{enumerate}
    \item We choose $N_n$ number of distinct bonds. This is done by generating a uniform random number between $r \in [0,1]$, which is then scaled by the number of bonds as follows $j = \mathrm{nint}\!\left(1 + r (N_m - 1)\right)$.  
        \item Assign the $j^\mathrm{th}$ site a polarity of $p_j=-1$.  
        \item Once the polarity of $N_n$ negative sites is assigned, we assign the polarity of the remaining sites $N_p$ to $p_j=+1$.
        \item The above step ensures the total number of polarity of negative and positive bonds, $N_n$ and $N_p$, respectively. 
    \end{enumerate}  

{\it Active Force Update:} We update the active force with the newly assigned polarity discussed in the previous steps. The active force on the $i$th monomer is given by, 
\begin{equation}
    \mathbf{F}_i^a =  f_a (p_{i-1} \hat \mathbf{t}_{i-1}+ p_{i}\hat \mathbf{t}_{i})/2,  
\end{equation}
        where $p_i$ denotes the polarity.  The active force on the terminal ends $F_1^a= f_a (p_1 \hat{\mathbf{t}}_1)/2  $ and $F_{N_m}^a= f_a (p_{N_m-1} \hat{\mathbf{t}}_{N_m-1})/2  $. 
      With this active force, particle positions are updated as given in Eq.~\ref{Eq:update}.   





\bibliography{Main}